\documentclass{emulateapj}

\newcommand{\kms}{\mbox{km\,s$^{-1}$}}

\shorttitle{Chemical Homogeneity in Collinder 261}
\shortauthors{De Silva et al.}

\begin{document}

\title{Chemical Homogeneity in Collinder 261\altaffilmark{*} and \\Implications for Chemical Tagging}


\author{G.M. De Silva\altaffilmark{1} and K.C. Freeman and M. Asplund}
\affil{Mount Stromlo Observatory, Australian National University,
Weston ACT 2611, Australia}
\email{gdesilva@eso.org}

\author{J. Bland-Hawthorn}
\affil{Anglo-Australian Observatory, Eastwood NSW 2122, Australia}
\email{jbh@aao.gov.au}

\author{M.S. Bessell}
\affil{Mount Stromlo Observatory, Australian National University,
Weston ACT 2611, Australia}

\author{R. Collet}
\affil{Department of Astronomy and Space Physics, Uppsala University, BOX 515, SE-751, Sweden} 
\altaffiltext{1}{Now at European Southern Observatory, Alonso de Cordova 3107, Casilla 19001, Santiago 19, Chile}
\altaffiltext{*}{Based on observations collected during ESO VLT-UT2 Programme 73.D-0716A at the European Southern Observatory, Paranal, Chile}



\begin{abstract}
This paper presents abundances for 12 red giants of the old open cluster Collinder 261 based on spectra from VLT/UVES. Abundances were derived 
  for Na, Mg, Si, Ca, Mn, Fe, Ni, Zr and Ba.  We find the cluster has
  a solar-level metallicity of [Fe/H] = -0.03 dex. However most $\alpha$
  and s-process elements were found to be enhanced. The
 star-to-star scatter was consistent with the expected measurement uncertainty
  for all elements.  The observed rms scatter is as follows: Na = 0.07, Mg = 0.05, Si = 0.06, Ca = 0.05, Mn = 0.03, Fe = 0.02, Ni = 0.04, Zr = 0.12, and Ba = 0.03 dex. The intrinsic scatter was estimated to be less than 0.05 dex. Such high levels of homogeneity indicate that chemical information remains preserved in this old open cluster.\\

We use the chemical homogeneity we have now established in Cr 261, Hyades and the HR1614 moving group to examine the uniqueness of the individual cluster abundance patterns, ie. chemical signatures. We demonstrate that the three studied clusters have unique chemical signatures, and discuss how other such signatures may be searched for in the future. Our findings support the prospect of chemically tagging disk stars to common formation sites in order to unravel the dissipative history of the Galactic disk.

\end{abstract}





\section{Introduction}

Old open clusters are rare fossils of the star formation history of
the Galactic disk. The majority of stars born in open clusters will
disperse into the Galaxy background within the first Gyr \citep{pjm}; the existence of several very old open clusters of ages
around 10 Gyr offer a unique opportunity to study the early evolution
of the disk. These important structures are not easily studied as they are rare and difficult to observe as most of them reside in the outer disk \citep{friel95}.  \\

Collinder 261 is an exception as it is located within the inner disk at a
Galactic radius of 7.5 kpc, with an estimated age range of 5 to 11 Gyr \citep{janes94,gozzoli,carraro}. This rich cluster has been previously
studied in the literature with membership established by photometric and
radial velocity studies \citep{friel02}. Several high resolution studies have also targeted Collinder 261. Most recently \citet[hereafter C05]{c05} presented a high resolution abundance analysis of six giants. They estimated a mean
metallicity of [Fe/H] = -0.03 dex, and found Na, Mg, Si and Ba to be
enhanced. Previously \citet[hereafter F03]{friel03}, also using high resolution spectra, estimated a mean metallicity of -0.22 dex and also found Na, Al and Si to be enhanced. Both studies found the other elements to be at solar or sub-solar levels.\\

Given the old age of the cluster, the chemical information will 
provide us with the conditions of the protocluster gas cloud during
the early stage of the disk. The observed $\alpha$ enhancement for Cr~261 is a
consistent pattern observed in old open clusters (see Table 7 of F03). This is a sign of a rapid enrichment history, which is to be expected at the early stages of disk formation. Since the age of Cr 261 is comparable to the age of the disk, it is likely to have formed shortly after the disk began to form, therefore its chemical evolution must have been relatively quick. C05 find that Ba, which is thought to be produced predominantly in AGB stars (although a smaller r-process component may be produced from Type II SN \citep{pagel}), is also enhanced in Cr 261. The number of studies on the heavier n-capture elements is few for old open clusters. Work on other n-capture elements would be helpful in exploring the enrichment history of these elements. \\

C05 find the star-to-star scatter to be low $\langle\sigma\rangle \sim$ 0.08 dex, which is within their estimated abundance uncertainties. The indication
of chemical homogeneity in Cr~261 is important for testing the
viability of chemical tagging as proposed by \citet{fbh02}, however the difference in the estimated
metallicities between C05 and F03 indicated the need for an
independent abundance analysis. Further, both studies were based on a
small sample of stars. In our analysis of Cr 261 we have doubled their sample size in order to establish a firmer level of
homogeneity. If chemical homogeneity within the 0.05 dex level can be 
firmly established for Cr 261, this would imply that the chemical
signature laid down at birth has been preserved over the time
evolution of the cluster and is indeed a true tracer of star formation
history in the disk. With the aim of testing these ideas we proceed with our study on Cr 261.\\

\section{Observations}

Because Cr 261 is relatively distant, we chose to observe giants in this cluster. A total of 18 giant stars of Cr 261 were submitted for service mode
observations in May 2004 on the 8m VLT, making use of the UVES Red arm with the
FLAMES fibre array which allows up to 6 stars to be observed
simultaneously.  The UVES Red arm standard setting provides a spectral resolution of 47,000 and complete spectra from 4200\AA\ to 6200\AA. \\

The method of observing was such that for one
telescope pointing, three different fibre combinations were
executed, with six stars in each fibre combination. This is possible
because the open cluster members of interest are located within
the instrument field of view. Each
fibre configuration was observed for a total of 5 hours to obtain
the required signal to noise. In practice the 5 hours were broken into
several one hour observing blocks to facilitate the service observing
queue. Our restrictions on the observing conditions was that the
seeing be better than 1.2 arcsec and airmass no more than 1.2. \\

The final data set reduced with the UVES ESO-MIDAS pipeline consists of high quality spectra for 12 stars, with the 6 other stars having very little signal. Since the magnitudes of all stars were comparable, we assume that misalignment 
of a few fibres was the cause. The spectra of the 12 stars
have a S/N between 80 - 100, sufficient for our abundance
analysis. Table \ref{cr261sample} presents a summary of the
stars we have studied.\\

\section{Abundance Analysis}

\subsection{Model Atmospheres and Spectral Lines}
The abundance analysis makes use of the MOOG code
\citep{sneden73} for LTE EW analysis and spectral syntheses.
Initial analysis was undertaken with interpolated Kurucz model
atmospheres based on the ATLAS9 code \citep{cas97} with no convective
overshoot. Later, our abundances were re-evaluated using MARCS models \citep{asplund97}, primarily to check the accuracy of the Kurucz models for
the cooler stars, as well as to check for consistency in our abundance
analysis for the entire sample. \\

Abundances for a range of elements covering each of the $\alpha$, Fe-peak
and n-capture groups were attempted. The list of lines used in this analysis is given in Table \ref{tab:line}. The $gf$ values for the detected lines of Na, Mg, Al, Si, Ca, Ni, and Zr were obtained from a combination of lines from \citet{ap04,yong05,reddy03} and \citet{P03}. For Mn, the $gf$ values were taken from \citet{prochaska00} and include the effects of hyperfine splitting. The main sources of the Fe~{\sc i} line data is the laboratory measurements by the Oxford group
(Blackwell et al., 1979a,b, 1995 and references therein). This was supplemented
by additional lines from \citet{reddy03}. For Fe {\sc ii} we adopt the $gf$ values from \citet{biemont91,P03} and \citet{fe2lines}. Ba $gf$  values were adopted from \citet{mcw98}. Although abundance determinations were attempted, most of the heavier s- and r-process element abundances could not be accurately derived, especially for the cooler stars because blending of lines was too high to allow an accurate abundance estimate.\\

\subsection{Stellar parameters}

We derive the stellar parameters based on spectroscopy. Abundances for all Fe {\sc i} and {\sc ii} lines were computed from the measured EWs. T$_{eff}$ was derived by requiring excitation equilibrium. Microturbulence was derived from the condition that Fe {\sc i} lines show no trend with EW. Log g was derived via ionization equilibrium, ie. the abundances from Fe {\sc i} equals Fe {\sc ii}.  The resulting stellar parameters are given in Table \ref{cr261params}. We also compare our derived parameters with those derived in the literature for the stars we have in common. Our parameters are in better agreement with C05 than with F03. \\

\subsection{Elemental Abundances}

The abundances were derived by EW measurements or spectral synthesis
depending on the strength and level of blending. All $\alpha$, Fe-peak, and Zr abundances were estimated by EW measurements as their transitions lines
were sufficiently strong and unblended to accurately measure EWs. Inital Ba abundances were also obtained via EW measurements, not taking into account any hyperfine structures (HFS). Later we carried out spectral synthesis of the Ba lines, incorporating the HFS given by \citet{mcw98} assuming a solar isotopic ratio. By taking into account HFS, we find the Ba abundance drops by about 0.15 dex. This later Ba abundances are adopted throught this paper.\\

 The abundance derivation of the heavier s- and r-process elements (eg. Nd,
Eu) were attempted by spectral synthesis. Although spectral synthesis
allows for abundance derivation from some blended lines, the spectral
regions of these lines were far too blended with many other unidentified lines also present. Our synthetic input line list
was primarily composed of spectral line data as provided by the VALD
database. All known element line data within the specific wavelength region of our lines of interest were extracted from the VALD database to suit the
stellar parameters. However fits to the observed spectra were
poor, likely due to inaccurate and incomplete atomic line list. As a result, we were unable to derive accurate abundances for the heavier s- and r-process elements. We note that C05 also did not obtain abundances for elements heavier than Ba. \\ 

Since our aim is to determine the level of homogeneity within the
cluster, we derive abundances with reference to the cluster star
2307, as it has an effective temperature in the middle of the range for our sample stars. The final differential
abundances ($\Delta$[X/H]), were derived by subtracting the absolute
abundance of each individual line of the reference star from the same
line of the sample stars and taking the mean for each element. The
advantage of such relative abundances is that the uncertainty due to
systematic errors (eg. errors in $gf$ values) are much reduced. Our
differential abundances are plotted in Figure \ref{cr261fe} for Fe,
and Figure \ref{cr261light} for elements from Na to Ba. We present our absolute abundances in log $\epsilon$ form in Table \ref{ab_table}.\\

The abundances for star 2311 are higher in all elements and deviate significantly from the other cluster member abundances. This star is represented by an open circle in Figures \ref{cr261fe} and \ref{cr261light}. A radial velocity analysis performed at a later stage shows that this star is a non-member. We will further discuss this in Section \ref{ch6:rv}. \\

\subsection{Error Analysis}\label{c6:errors}

The main sources of errors are the error associated with EW measurements, continuum placement and stellar parameters, as well as the number of lines used to calculate the final abundance. Errors in the atomic line data and model atmospheres are least likely to affect the estimated levels of chemical
homogeneity as we are employing a differential abundance analysis relative to a cluster member. \\

Abundance dependencies on the stellar parameters and EW measurements, as well as the typical values of the total estimated uncertainty for each element are given in Table \ref{tab:error}. The error in EWs estimated by repeated measurements of each line, is between 2m\AA\ to 10m\AA\, depending on the strength of the lines. The typical error in the stellar parameters are around $\delta T_{eff}$ = 50 K, $\delta$log \emph{g} = 0.1 cm\,s$^{-2}$ and $\delta \xi$ = 0.2\kms. \\

Our analysis is based on Kurucz models. However due to the cooler T$_{eff}$ for some of the sample stars, we tested our
results using MARCS models for three stars (2285, 2288, 3709) which cover the full temperature range. For the hotter star 2285 the
change in abundance was minimal for all elements with a mean difference of $\pm$ 0.01 dex. For star 2288, differences of 0.07 and 0.1 for Si and Ni were found. For the coolest star 3709, larger differences of 0.15 dex for Na and Ca, and
0.35 dex for Zr were found. Table \ref{models1} summarizes these differences. These results were based on the same stellar parameters derived initially with Kurucz models. To enable a better comparison, the stellar microturbulence was then adjusted by 0.2 \kms\ to fit the MARCS models. This resulted in a better agreement with our initial results, with a mean difference of about $\pm$ 0.03 dex. A summary of the latter results are presented in Table \ref{models2}.\\

\section{Radial velocities}
\label{ch6:rv}

We calculated the radial velocities of the sample stars to check for any possible non-members. The RVs were determined by Fourier transform cross
correlation of template spectra with observed spectra, making use of the IRAF packages RVSAO/XCSAO \citep{rvsao,xcsao}. From the available spectra and template wavelength range, RVs were estimated using the blue region from 4200 - 4400 \AA. Template spectra from \citet{zwitter} were obtained via private communication from M.
Williams. Since the stellar parameters were already established from our
earlier spectroscopic studies, templates matching closest to the sample
parameters were selected for the cross correlation. Our errors are within 2 \kms. \\

Table \ref{cr261rv} shows our derived heliocentric RVs, as well as those obtained by \citet{friel02}. Our results are on average higher than \citet{friel02} by 5 \kms; larger differences are seen for the two stars 2311 and 3029. \citet{friel02} find star 2311 to have a RV of -30 \kms\ similar to their derived cluster mean value. This is inconsistent with our result of -18 \kms; our velocity indicates that star 2311 is likely to be a non-member of the cluster. Conversely, \citet{friel02} find the star 3029 to have a RV of -16 \kms, although they did not class it a non-member. Our results show that 3029 has a RV of -24 \kms\ , which places it well within the cluster RV range.\\

\section{Discussion}

\subsection{Comparison with Carretta et al. (2005)}

Our results based on a larger sample of stars are comparable to the
mean abundances found earlier by C05, although differences are present
in the individual stars. Five out of six of their stars are in
common with our study, and in Figure \ref{cr261comp} we compare the abundances of these five individual stars. The largest difference seen for the
coolest star 3709, which is at the tip of the red giant branch, is
interesting. The adopted stellar parameters are very similar in both
studies, and are therefore unlikely to be the reason for the abundance
difference. However the choice of Kurucz vs.~MARCS model atmospheres, as well as  other differences in methodology may contribute to the abundance difference.
Our abundance estimates gives star 3709 similar abundances to the rest of the cluster, while C05 found this star to deviate from their sample, and it was discarded when determining their cluster mean abundances. \\

Mean cluster abundances relative to solar are presented in column 5 of Table \ref{tab:tracks}. On average, we find that Na, Mg and Si are enhanced by about 0.15 dex relative to the Sun. Ca on the other hand was found to be at solar levels. The Mn and Ni abundances are slightly below solar, but generally follow Fe. We also measured Zr abundances and find the mean value to be below solar. However there is considerable scatter. These results follow the abundance patterns found by C05, but we find that the enrichment levels are not as high. On average C05 find $\alpha$ enhancement to $\approx$ 0.25 dex levels, while we observe enhancement to $\approx$ 0.15 dex levels. The Ba abundances we have derived are significantly lower than that of C05. C05 find the mean Ba enhancement to be 0.3 dex. Our results, based on spectral synthesis and incorporating HFS, show the Ba abundance in Cr 261 is close to solar. C05 did not take into account any HFS and this is likely to be the main reason for the large abundance difference. Our inital Ba measurements based on EW measurments and not taking into account the effects of HFS, resulted a mean Ba enhancement of $\approx$ 0.15 dex. Although this is still lower than the measurements of C05, the difference is comparable with the differences seen for the other $\alpha$ elements between the two studies.\\

The enhanced $\alpha$ abundances indicate that the contribution of
material from AGB and Type II SN is greater than that of Type Ia SN,
where most of the Fe-peak elements are thought to be formed. 
It would also be interesting to check if some of the enhancements (e.g. for Na) in Cr 261 may be linked to internal mixing (e.g. from Ne-Na burning chain) in the sample giants, by measuring the abundances in dwarf stars. If so these abundances may not be representative of the proto-cluster cloud. It is thought that such processes play a larger role in globular clusters \citep{gratton04}, and have not been previously observed in open clusters.\\

\subsection{Chemical Homogeneity}\label{c6:homo}

Our results indicate that Collinder 261 is chemically homogeneous. Disregarding Zr where the abundance uncertainty was large, we find the mean star-to-star scatter across a range of elements to be $\approx$ 0.05 dex. The observed rms scatter $\sigma_{obs}$ is summarized in Table \ref{cr261scatter}. In all cases the observed scatter is within the expected uncertainty in the abundances. This implies that the low intrinsic scatter in this old cluster is undetectable at the current accuracy. C05 also did not find any significant scatter across their sample of 5 stars, although their scatter increases if they included star 3709 which we believe to be a member. Taking into account possible uncertainty in our error analysis, we use the smallest plausible estimate of our measurement errors in order to derive an upper limit on the intrinsic scatter, which is also presented in Table \ref{cr261scatter}. As described in \citet{hr1614}, we derive the confidence interval for the upper limit on the intrinsic scatter, taking into account the sampling error on the observed scatter and a 10\% uncertainty on the adopted measurement errors. We find the upper limits given in Table \ref{cr261scatter} are approximately 90\% confidence limits for the intrinsic scatter in all elements except Si and Zr, whose upper limits are approximately 80\% confidence limits. Given zero intrinsic scatter and our measurement errors, the probability of obtaining the observed scatter was also calculated based on a $\chi^{2}$ analysis. Figure \ref{probcr261} shows this probability for the studied elements.\\

The level of homogeneity seen in this cluster implies that the original abundances remain preserved in stars despite their stellar evolution, and pollution has no detectable effect for these elements. In making the above calculations of the intrinsic scatter, we have omitted the star 2311, which was found to be enriched compared to the cluster mean. Although \citet{friel02} find this star to be a cluster member, our RV analysis indicates that it is very likely a non-member of Cr 261. The ability to chemically distinguish such non-members without the prior dynamical information, as was the case for star 2311, is an encouraging demonstration for the prospect of future chemical tagging.

\section{Implications for Chemical Tagging}

\subsection{Chemical Tagging}\label{c7:chemtag}

A major goal of near-field cosmology is to tag or associate individual stars with elements of the proto-cloud. Since the Galactic disk formed dissipatively and evolved dynamically, much of the dynamical information is lost. Any dynamical probing of the disk will only provide insights back to the epoch of last dynamical scattering. However, the chemical information within the stars survived the disk's dissipative history. In order to follow the sequence of events involved in dissipation, the critical components which need to be re-assembled are the ancient individual star-forming aggregates in the disk. If star-forming aggregates can preserve unique \emph{chemical signatures} within their member stars, we can use these
signatures to tag dispersed individual stars to a common formation event. With sufficiently detailed abundances we would be able to reconstruct the stellar aggregates which have long since diffused into the Galaxy background \citep{fbh02}.\\

As discussed by \citet{bhf04PASA}, there are some basic requirements for the feasibility of such large scale chemical reconstruction. For example, the stellar chemical abundances must reflect the composition of the parent cloud for certain key elements. Further these key elemental abundances must not be rigidly coupled, and have sufficient abundance dispersions to allow for identification of unique sites of formation within the large chemical inventory. Identifying these suitable key elements for chemical tagging are also part of the viability tests. Over the past decades evidence has gathered for a large dispersion in elemental abundances, particularly for the heavier n-capture elements at low metallicities. Based on several recent Galactic surveys \citep[eg.][]{ap04,bensby, reddy03}, we can approximate the mean scatter to about 0.2 dex for all studied elements, over a metallicity range of about -1.0 $<$ [Fe/H] $<$ +0.3 dex. With such a range in the field, individual abundance measurements to 0.05 dex level accuracies provide us with four distinguishable abundance levels. Therefore, how many decoupled elements will be needed to identify the unique chemical signatures of the disk within the larger chemical inventory? \citet{fbh02} and \citet{bhf04PASA} show the number of unique star-formation sites over the entire disk is between 10$^{6}$ to 10$^{8}$. Assuming only four distinguishable abundance levels, then the identification of 10 to 15 decoupled elements will provide the required number (ie. 4$^{15}$) of independent cells in chemical space.\\
 
Another basic requirement for chemical tagging is chemical homogeneity within present day stellar aggregates in the disk, such as open clusters and moving groups. \citet{conti65} was perhaps the first to attempt to quantify the level of homogeneity in open clusters. Although high resolution abundance studies on open clusters are limited, some recently published measurements of both light and heavy element abundances in open clusters demonstrate chemical homogeneity, albeit for only a few stars, and lend support to the prospect of chemical tagging (eg. \citet{ford}, \citet{schuler03}, \citet{gonzalez2000}, \citet{tau00}). Our study on the chemical homogeneity in the Hyades in \citet[Paper I]{desilva} and the HR1614 moving group in \citet[Paper II]{hr1614}, as well as in Cr 261 presented here, show chemical homogeneity over a range of $\alpha$, Fe-peak, and heavy elements for larger sample sets, satisfying the primary requirements for the chemical tagging technique. Such demonstrations of highly chemically homogeneous star clusters has opened the possibilities for the next set of tests for the viability of chemical tagging. The next task is to identify the uniqueness of the individual cluster abundance patterns, ie. chemical signatures. The results presented earlier in this paper, as well as in Paper I and Paper II, can be adapted to obtain a preliminary understanding of the unique chemical signatures, and how such signatures may be searched for in the future. \\

\subsection{Chemical Signatures}\label{c7:signature}

A summary of our results for the three clusters are presented in Table
\ref{tab:tracks}. The abundances are presented relative to solar. We have adopted the solar photospheric values from \citet{grev98} for all elements except Fe \citep{snedenfe}, Nd \citep{dlsc03} and Eu \citep{lawler01}. We present Table \ref{tab:tracks} graphically in Figure \ref{track}, and in Figure \ref{trackfe} we replot the abundances relative to Fe. The different colored shapes represent the different clusters. Each data point represents the cluster mean value and the
error bars indicate the observed scatter. For the Hyades, the
abundances of Fe and $\alpha$ elements were adopted from
\citet{P03}, while the heavier element abundances are from Paper I. Note that the non-members or other peculiar stars found in the earlier studies have not been included in determining the mean abundance values for any of the clusters.\\

Figure \ref{track} shows that the three clusters clearly have their own
chemical abundance pattern, with little overlap. The Hyades abundances
follow a slightly super-solar abundance pattern for most elements,
however Mg is underabundant and Ba is greatly enhanced. The HR1614 group abundances follow a super-solar abundance level of about 0.25 dex for most elements, Zr, Ce, and Nd are only enhanced by about 0.15 dex, and Ba is again highly enhanced. The r-process element Eu is also enhanced by 0.21 dex. Collinder 261 follows a different pattern with Na, Mg and Si enhanced to about 0.15 dex level, Ca, Fe and Ba abundances at solar-levels, and Mn, Ni and Zr abundances are slightly below solar at -0.04 dex. In summary, the abundance signatures observed for our sample clusters are all different. \\

The large separations seen in Figure \ref{track} due to the difference in mean metallicity can be removed by plotting the abundances relative to Fe (Figure \ref{trackfe}). The trends in [X/Fe] highlights the other signatures of these clusters besides the principal Fe component. The $\alpha$ elements seem to be the next dominant component, followed by Ba, and the Fe-peak elements. More elements would provide a better understanding of the subsequent components. For a larger sample of clusters we expect that a Principle Component Analysis (PCA) would allow us to identify the major elements or element groups required to uniquely identify a cluster signature.\\

\subsection{Comparison to disk abundances}

It is of interest to compare our cluster abundance signatures with the abundance trends of field stars in the Galactic disk. Cepheids are excellent stellar objects to study the recent state of the young disk at different Galactocentric radii. To examine any possible differences within the solar neighborhood, we plot our cluster abundances and the abundances of cepheids by \citet{cepheids} against Galactocentric radius in Figures \ref{ceph_fe} to \ref{ceph_heavy}. The Hyades and Collinder 261 reside close to the Sun, within 1 kpc. Adopting a site for the HR1614 moving group is difficult since the member stars are dispersed throughout the Galaxy; since our studied stars are located near the Sun, we have adopted the solar radius.\\

In Figure \ref{ceph_fe}, we see that Cr 261 lies within the range of cepheid metallicities. Since Cr 261 is an old open cluster of age $\sim$ 8 Gyr, the agreement with the cepheid metallicities indicate that the chemical evolution in this region of the disk proceeded quickly and then remained relatively quiet. The Hyades is close to the upper limit of the cepheid metallicity. A clear deviation is seen for the HR1614 moving group compared to the cepheid abundances at the solar radius. This may be due to the birth site of the group in the inner disk, where a few cepheids are also observed to have high metallicities at R$_{GC} \sim$ 6.5 kpc. For the other elements, most are in good agreement with the cepheids. The enhanced Na and depleted Mg in the cepheids may conceivably be due to the internal Ne-Na and Mg-Al cycles. Significant deviations are also seen for Si, where Cr 261 is overabundant and the Hyades is underabundant in comparison to the cepheids. These deviations may be local inhomogeneities that distinguish the individual clusters. Finding such deviations is of great interest for chemical tagging as it highlights again the uniqueness of the signatures.\\

To check for any further deviations of our cluster signatures, we next compare our results with the abundance patterns of disk stars. Figure \ref{field} shows our cluster abundances overplotted with the results by \citet{reddy03}, \citet{ed93}, and \citet{ap04} for the common elements in nearby disk stars. The sample of \citet{ap04} includes all stars more luminous than M$_{V}$ = 6.5 mag within 14.5 pc from the Sun, while the samples of \citet{reddy03} and \citet{ed93} contains F and G dwarf stars chosen to be evenly distributed over a metallicity range of about -1.0 $<$ [Me/H] $<$ +0.3 dex. The sample of \citet{reddy03} contains almost exclusively thin disk stars, while the sample of \citet{ed93} contains both thin and thick disk stars. \\

In Figure \ref{field} we are now focusing on the local stellar populations, mostly within 1 kpc of the Sun. In comparison to the field stars, one notices that some element abundances of the open clusters do not match the abundance range of the field. Mg and Si is underabundant in the Hyades in comparison to the field, while Ca is similar and Ba is greatly enhanced. Collinder 261 lies within the field abundances, although the abundances of Mg and Si are close to the upper limit. The HR1614 moving group is among the most metal-rich stars in the field. Its abundances are mostly within the field, except for Ba which is enhanced. However not many field stars are studied at such high metallicities. These deviations from the field are an indication of the uniqueness of the individual cluster signatures and is likely related to the different chemical enrichment history experienced by the clusters compared to the solar neighborhood. \\

Overall we see that the chemical signature of the clusters do not necessarily match the nearby young and old disk stars, particularly for Na, Mg, Si and Ba which are the most deviant elements. In general it seems the light $\alpha$ and heavy s-process elements are deviant, while the heavier $\alpha$ (Ca) and light s-process (Zr) elements follow the field star profiles. It may be that localized inhomogeneities at the time and site of formation of the clusters are linked to the synthesis processes of these elements, which would give rise to the observed variations. In the following section we will discuss the known formation processes of the individual elements.\\

\subsection{Other open clusters}
Finally, we compare our results with abundances of other open clusters
in the literature, treating each element individually. The number of
clusters which have been subject to high resolution abundance analysis
is small. Even smaller is the number of clusters with abundances
derived for a range of elements (from light to n-capture elements). We
combine our results with literature abundances for the open clusters
Be 20, Be 29, Be 31, and NGC 2141 \citep{yong05}, M11
\citep{gonzalez2000}, Tombaugh 2 and Mellote 71 \citep{brown}, NGC
2243 and Mellote 66 \citep{gratton}, NGC 7789 and M67
\citep[respectively]{tau05,tau00}, and NGC 6819 \citep{ngc6819}.\\ 

When comparing abundances from various sources, one must note the
possible systematic effects arising as a result of different
methodologies, such as differences in the adopted solar values, the
$gf$ values of the atomic lines, etc. Where an inverted solar
analysis was performed the abundances were taken as published, but for
studies where different solar values were adopted we have recalculated
the abundances relative to our adopted solar values to enable a better
comparison. Figures \ref{ocna} to \ref{oceu} show the resulting
plots. Various colors represent different authors and symbols denote
the different clusters. All cluster mean abundances are plotted. The
clusters cover a large range in age, metallicity and Galactocentric
radii. We can use these plots to examine the decoupled nature of the
elements or element groups, and identify the dimensionality of the
chemical space within the studied elements. Some elements are
tightly locked to Fe, some show trends and significant scatter. \\

Of the $\alpha$ elements, Mg shows the largest scatter as well as a
decreasing trend with metallicity, while Si and Ca follow Fe with some
scatter. This may be an indication of substructure within the $\alpha$
element abundances. These elements are currently believed to be formed during
Type II SN and not modified internally in the lower mass
stars. However for massive stars if internal mixing is effective, the
Mg-Al cycle can alter the Mg abundances. An anti-correlation with Al
abundances would be a sign of such processes. Otherwise, enhanced Mg
and other $\alpha$ element abundances relative to Fe are likely
representative of a high star formation rate, where Type II SN
dominate over the Type Ia SN. As to why only Mg shows a larger scatter
than other $\alpha$ elements remains a question; it is likely due to a different synthesis process that has not been fully understood. Note that Si and Ca are the two pure $\alpha$ elements formed only via the alpha-capture process; other processes are involved in the evolution of Mg (as well as Ti) abundances.\\

Ni and Mn belong to the so-called Fe-peak elements. They are thought
to be the products of Type Ia SN. In theory these elemental abundances
should be tightly coupled to Fe. While Ni is tightly correlated to Fe
in Figure \ref{ocmn}, Mn shows a clear increasing trend with
metallicity. This difference is interesting as it demonstrates the
decoupled nature of abundances {\it within} the Fe-peak elements. It
may be that there are other sources for Mn synthesis besides Type Ia
SN which causes the decoupled nature from other Fe-peak
elements. For example, {\citet{mcwil03} suggest that Type II SN is also a
contributor of Mn, and that both Type Ia and Type II SN yields are metallicity
dependent \citep[see also][]{shetrone}. Such an explanation better
fit the current observations.  \\

In the open cluster abundances, much scatter is seen in Na, Zr, Ba and
Eu abundances. Na is thought to be synthesized in giant stars via the
Ne-Na chain and may result in self enrichment if convection is strong in
the studied stars. However in stars where internal mixing is not
significant we cannot expect such enrichment to take place. Na is more
likely to enter the ISM via Type II SN or in the ejecta of AGB stars. 
In that case, the enhancement in Na relative to Fe is consistent with
a rapid star formation rate. \\

Zr and Ba are s-process elements which are synthesized under a low
neutron flux environment such as in AGB stars, and are then expelled
via stellar ejecta. Zr is a light s-process element, while Ba is a
heavy s-process element, where the abundances of the heavy s-process
is consistently greater than the light s-process, demonstrating the
decoupled nature of the two groups. The scatter in their
abundances is also seen in the field stars (see Figure
\ref{field}). Whether this reflects a true scatter is
uncertain. Inconsistent measurement of abundances 
(eg. not including possible NLTE \citep{asplund05} and/or HFS effects) may also give
rise to the presently observed scatter. If the scatter is real, then
this implies several sources for element synthesis. For example, Ba may
be synthesized by the r-process via Type II SN in the older stars at
low metallicities \citep{pagel}. \\

Finally, Eu is thought to be produced by the r-process during Type II
SN, in a high neutron flux environment. The general slope seen in
Figure \ref{oceu} is consistent with enrichment by Type II SN, however
exceptions are Mel66, NGC 2343 and Tom2. Unfortunately, no other
r-process elements have been studied to confirm the lower abundances
for these clusters. As for the $\alpha$ elements, enhanced Eu to Fe
ratios is an indication of a period of rapid chemical evolution. \\

In summary, we see that several decoupled elements show abundance
dispersions which play a significant role in defining a large chemical abundance space, which \citet{fbh02} refer to as ${\cal C}$-space. A PCA
analysis is likely to reveal more specific details about the dominant
components of the ${\cal C}$-space, once larger samples of clusters
covering a large range of elements become available. This preliminary
examination shows that at least the $\alpha$, Fe-peak, light and heavy
s-process, and r-process elements, with particular attention to Na,
Mg, Si, Mn, Zr, Ba and Eu abundances in low mass stars, provide a starting
point for detecting chemical substructures. Following our present
discussion, if we consider the abundance trends of Mg from the other
$\alpha$ element abundances, as well as Mn abundances from the other
Fe-peak abundance patterns, we have at present a total of 8
decoupled groups already apparent in our study which was limited to 12
elements. It is conceivable that by exploring more elements, the required number of decoupled elements or element groups can be established, making the technique of chemical tagging more viable.\\

As discussed by \citet{fbh02} and \citet{bhf04PASA}, for large scale chemical tagging, accurate chemical and dynamical data will be required. Several major
current and planned Galactic surveys, as well as the availability of multi-fibre spectrographs on future ELTs will provide the required data. The tests we have
performed are a vital starting point for exploiting the detection of
disk substructure from the future data. Once the technique is well
tested and proven, chemical tagging will pave the way to obtain a
detailed physical picture of events that led to the formation of the
Galactic disk.

\acknowledgments
We thank Eileen Friel for kindly providing the co-ordinates of the Cr 261 stellar sample, Mary Williams for her assistance with determining radial velocities, and the anonymous referee for his useful suggestions. This research has made use of the Vienna Atomic Line Database (VALD), operated at Vienna, Austria, and the IRAF package distributed by the National Optical Astronomy Observatory, which is operated by the Association of Universities for Research in Astronomy, Inc., under cooperative agreement with the National Science Foundation.



\clearpage

\begin{thebibliography}{43}
\expandafter\ifx\csname natexlab\endcsname\relax\def\natexlab#1{#1}\fi

\bibitem[{{Allende Prieto} {et~al.}(2002){Allende Prieto}, {Asplund},
  {L{\'o}pez}, \& {Lambert}}]{fe2lines}
{Allende Prieto}, C., {Asplund}, M., {L{\'o}pez}, R.~J.~G., \& {Lambert}, D.~L.
  2002, \apj, 567, 544

\bibitem[{{Allende Prieto} {et~al.}(2004){Allende Prieto}, {Barklem},
  {Lambert}, \& {Cunha}}]{ap04}
{Allende Prieto}, C., {Barklem}, P.~S., {Lambert}, D.~L., \& {Cunha}, K. 2004,
  \aap, 420, 183

\bibitem[{{Andrievsky} {et~al.}(2002){Andrievsky}, {Kovtyukh}, {Luck},
  {L{\'e}pine}, {Bersier}, {Maciel}, {Barbuy}, {Klochkova}, {Panchuk}, \&
  {Karpischek}}]{cepheids}
{Andrievsky}, S.~M., {Kovtyukh}, V.~V., {Luck}, R.~E., {L{\'e}pine}, J.~R.~D.,
  {Bersier}, D., {Maciel}, W.~J., {Barbuy}, B., {Klochkova}, V.~G., {Panchuk},
  V.~E., \& {Karpischek}, R.~U. 2002, \aap, 381, 32

\bibitem[{{Asplund}(2005)}]{asplund05}
{Asplund}, M. 2005, \araa, 43, 481

\bibitem[{{Asplund} {et~al.}(1997){Asplund}, {Gustafsson}, {Kiselman}, \&
  {Eriksson}}]{asplund97}
{Asplund}, M., {Gustafsson}, B., {Kiselman}, D., \& {Eriksson}, K. 1997, \aap,
  318, 521

\bibitem[{{Bensby} {et~al.}(2003){Bensby}, {Feltzing}, \&
  {Lundstr{\"o}m}}]{bensby}
{Bensby}, T., {Feltzing}, S., \& {Lundstr{\"o}m}, I. 2003, \aap, 410, 527

\bibitem[{{Biemont} {et~al.}(1991){Biemont}, {Baudoux}, {Kurucz}, {Ansbacher},
  \& {Pinnington}}]{biemont91}
{Biemont}, E., {Baudoux}, M., {Kurucz}, R.~L., {Ansbacher}, W., \&
  {Pinnington}, E.~H. 1991, \aap, 249, 539

\bibitem[{{Bland-Hawthorn} \& {Freeman}(2004)}]{bhf04PASA}
{Bland-Hawthorn}, J., \& {Freeman}, K.~C. 2004, Publications of the
  Astronomical Society of Australia, 21, 110

\bibitem[{{Bragaglia} {et~al.}(2001){Bragaglia}, {Carretta}, {Gratton}, {Tosi},
  {Bonanno}, {Bruno}, {Cal{\`i}}, {Claudi}, {Cosentino}, {Desidera},
  {Farisato}, {Rebeschini}, \& {Scuderi}}]{ngc6819}
{Bragaglia}, A., {Carretta}, E., {Gratton}, R.~G., {Tosi}, M., {Bonanno}, G.,
  {Bruno}, P., {Cal{\`i}}, A., {Claudi}, R., {Cosentino}, R., {Desidera}, S.,
  {Farisato}, G., {Rebeschini}, M., \& {Scuderi}, S. 2001, \aj, 121, 327

\bibitem[{{Brown} {et~al.}(1996){Brown}, {Wallerstein}, {Geisler}, \&
  {Oke}}]{brown}
{Brown}, J.~A., {Wallerstein}, G., {Geisler}, D., \& {Oke}, J.~B. 1996, \aj,
  112, 1551

\bibitem[{{Carraro} {et~al.}(1998){Carraro}, {Ng}, \& {Portinari}}]{carraro}
{Carraro}, G., {Ng}, Y.~K., \& {Portinari}, L. 1998, \mnras, 296, 1045

\bibitem[{{Carretta} {et~al.}(2005){Carretta}, {Bragaglia}, {Gratton}, \&
  {Tosi}}]{c05}
{Carretta}, E., {Bragaglia}, A., {Gratton}, R.~G., \& {Tosi}, M. 2005, \aap,
  441, 131

\bibitem[{{Castelli} {et~al.}(1997){Castelli}, {Gratton}, \& {Kurucz}}]{cas97}
{Castelli}, F., {Gratton}, R.~G., \& {Kurucz}, R.~L. 1997, \aap, 318, 841

\bibitem[{{Conti} {et~al.}(1965){Conti}, {Wallerstein}, \& {Wing}}]{conti65}
{Conti}, P.~S., {Wallerstein}, G., \& {Wing}, R.~F. 1965, \apj, 142, 999

\bibitem[{{De Silva} {et~al.}(in press){De Silva}, {Freeman},
  {Bland-Hawthorn}, {Asplund}, \& {Bessell}}]{hr1614}
{De Silva}, G.~M., {Freeman}, K.~C., {Bland-Hawthorn}, J., {Asplund}, M., \&
  {Bessell}, M.~S., \aj,  in press (astro-ph/0610041)

\bibitem[{{De Silva} {et~al.}(2006){De Silva}, {Sneden}, {Paulson}, {Asplund},
  {Bland-Hawthorn}, {Bessell}, \& {Freeman}}]{desilva}
{De Silva}, G.~M., {Sneden}, C., {Paulson}, D.~B., {Asplund}, M.,
  {Bland-Hawthorn}, J., {Bessell}, M.~S., \& {Freeman}, K.~C. 2006, \aj, 131,
  455

\bibitem[{{Den Hartog} {et~al.}(2003){Den Hartog}, {Lawler}, {Sneden}, \&
  {Cowan}}]{dlsc03}
{Den Hartog}, E.~A., {Lawler}, J.~E., {Sneden}, C., \& {Cowan}, J.~J. 2003,
  \apjs, 148, 543

\bibitem[{{Edvardsson} {et~al.}(1993){Edvardsson}, {Andersen}, {Gustafsson},
  {Lambert}, {Nissen}, \& {Tomkin}}]{ed93}
{Edvardsson}, B., {Andersen}, J., {Gustafsson}, B., {Lambert}, D.~L., {Nissen},
  P.~E., \& {Tomkin}, J. 1993, \aap, 275, 101

\bibitem[{{Ford} {et~al.}(2005){Ford}, {Jeffries}, \& {Smalley}}]{ford}
{Ford}, A., {Jeffries}, R.~D., \& {Smalley}, B. 2005, \mnras, 364, 272

\bibitem[{{Freeman} \& {Bland-Hawthorn}(2002)}]{fbh02}
{Freeman}, K., \& {Bland-Hawthorn}, J. 2002, \araa, 40, 487

\bibitem[{{Friel}(1995)}]{friel95}
{Friel}, E.~D. 1995, \araa, 33, 381

\bibitem[{{Friel} {et~al.}(2003){Friel}, {Jacobson}, {Barrett}, {Fullton},
  {Balachandran}, \& {Pilachowski}}]{friel03}
{Friel}, E.~D., {Jacobson}, H.~R., {Barrett}, E., {Fullton}, L.,
  {Balachandran}, S.~C., \& {Pilachowski}, C.~A. 2003, \aj, 126, 2372

\bibitem[{{Friel} {et~al.}(2002){Friel}, {Janes}, {Tavarez}, {Scott},
  {Katsanis}, {Lotz}, {Hong}, \& {Miller}}]{friel02}
{Friel}, E.~D., {Janes}, K.~A., {Tavarez}, M., {Scott}, J., {Katsanis}, R.,
  {Lotz}, J., {Hong}, L., \& {Miller}, N. 2002, \aj, 124, 2693

\bibitem[{{Gonzalez} \& {Wallerstein}(2000)}]{gonzalez2000}
{Gonzalez}, G., \& {Wallerstein}, G. 2000, \pasp, 112, 1081

\bibitem[{{Gozzoli} {et~al.}(1996){Gozzoli}, {Tosi}, {Marconi}, \&
  {Bragaglia}}]{gozzoli}
{Gozzoli}, E., {Tosi}, M., {Marconi}, G., \& {Bragaglia}, A. 1996, \mnras, 283,
  66

\bibitem[{{Gratton} {et~al.}(2004){Gratton}, {Sneden}, \&
  {Carretta}}]{gratton04}
{Gratton}, R., {Sneden}, C., \& {Carretta}, E. 2004, \araa, 42, 385

\bibitem[{{Gratton} \& {Contarini}(1994)}]{gratton}
{Gratton}, R.~G., \& {Contarini}, G. 1994, \aap, 283, 911

\bibitem[{{Grevesse} \& {Sauval}(1998)}]{grev98}
{Grevesse}, N., \& {Sauval}, A.~J. 1998, Space Science Reviews, 85, 161

\bibitem[{{Janes} \& {Phelps}(1994)}]{janes94}
{Janes}, K.~A., \& {Phelps}, R.~L. 1994, \aj, 108, 1773

\bibitem[{{Kurtz} \& {Mink}(1998)}]{rvsao}
{Kurtz}, M.~J., \& {Mink}, D.~J. 1998, \pasp, 110, 934

\bibitem[{{Kurtz} {et~al.}(1992){Kurtz}, {Mink}, {Wyatt}, {Fabricant},
  {Torres}, {Kriss}, \& {Tonry}}]{xcsao}
{Kurtz}, M.~J., {Mink}, D.~J., {Wyatt}, W.~F., {Fabricant}, D.~G., {Torres},
  G., {Kriss}, G.~A., \& {Tonry}, J.~L. 1992, in ASP Conf. Ser. 25:
  Astronomical Data Analysis Software and Systems I, 432--+

\bibitem[{{Lawler} {et~al.}(2001){Lawler}, {Wickliffe}, {den Hartog}, \&
  {Sneden}}]{lawler01}
{Lawler}, J.~E., {Wickliffe}, M.~E., {den Hartog}, E.~A., \& {Sneden}, C. 2001,
  \apj, 563, 1075

\bibitem[{{McWilliam}(1998){McWilliam}}]{mcw98}
{McWilliam}, A. 1998, \apj, 115, 1640

\bibitem[{{McWilliam} {et~al.}(2003){McWilliam}, {Rich}, \&
  {Smecker-Hane}}]{mcwil03}
{McWilliam}, A., {Rich}, R.~M., \& {Smecker-Hane}, T.~A. 2003, \apjl, 592, L21

\bibitem[{{Pagel} \& {Tautvaisiene}(1997)}]{pagel}
{Pagel}, B.~E.~J., \& {Tautvaisiene}, G. 1997, \mnras, 288, 108

\bibitem[{{Paulson} {et~al.}(2003){Paulson}, {Sneden}, \& {Cochran}}]{P03}
{Paulson}, D.~B., {Sneden}, C., \& {Cochran}, W.~D. 2003, \aj, 125, 3185

\bibitem[{{Phelps} {et~al.}(1994){Phelps}, {Janes}, \& {Montgomery}}]{pjm}
{Phelps}, R.~L., {Janes}, K.~A., \& {Montgomery}, K.~A. 1994, \aj, 107, 1079

\bibitem[{{Prochaska} \& {McWilliam}(2000)}]{prochaska00}
{Prochaska}, J.~X., \& {McWilliam}, A. 2000, \apjl, 537, L57

\bibitem[{{Prochaska} {et~al.}(2000){Prochaska}, {Naumov}, {Carney},
  {McWilliam}, \& {Wolfe}}]{prochaskaba}
{Prochaska}, J.~X., {Naumov}, S.~O., {Carney}, B.~W., {McWilliam}, A., \&
  {Wolfe}, A.~M. 2000, \aj, 120, 2513

\bibitem[{{Reddy} {et~al.}(2003){Reddy}, {Tomkin}, {Lambert}, \& {Allende
  Prieto}}]{reddy03}
{Reddy}, B.~E., {Tomkin}, J., {Lambert}, D.~L., \& {Allende Prieto}, C. 2003,
  \mnras, 340, 304

\bibitem[{{Schuler} {et~al.}(2003){Schuler}, {King}, {Fischer}, {Soderblom}, \&
  {Jones}}]{schuler03}
{Schuler}, S.~C., {King}, J.~R., {Fischer}, D.~A., {Soderblom}, D.~R., \&
  {Jones}, B.~F. 2003, \aj, 125, 2085

\bibitem[{{Shetrone} {et~al.}(2003){Shetrone}, {Venn}, {Tolstoy}, {Primas},
  {Hill}, \& {Kaufer}}]{shetrone}
{Shetrone}, M., {Venn}, K.~A., {Tolstoy}, E., {Primas}, F., {Hill}, V., \&
  {Kaufer}, A. 2003, \aj, 125, 684

\bibitem[{{Sneden} {et~al.}(1992){Sneden}, {Kraft}, {Prosser}, \&
  {Langer}}]{snedenfe}
{Sneden}, C., {Kraft}, R.~P., {Prosser}, C.~F., \& {Langer}, G.~E. 1992, \aj,
  104, 2121

\bibitem[{{Sneden}(1973)}]{sneden73}
{Sneden}, C.~A. 1973, Ph.D.~Thesis

\bibitem[{{Tautvai{\v s}ien{\.e}} {et~al.}(2005){Tautvai{\v s}ien{\.e}},
  {Edvardsson}, {Puzeras}, \& {Ilyin}}]{tau05}
{Tautvai{\v s}ien{\.e}}, G., {Edvardsson}, B., {Puzeras}, E., \& {Ilyin}, I.
  2005, \aap, 431, 933

\bibitem[{{Tautvai{\v s}iene} {et~al.}(2000){Tautvai{\v s}iene}, {Edvardsson},
  {Tuominen}, \& {Ilyin}}]{tau00}
{Tautvai{\v s}iene}, G., {Edvardsson}, B., {Tuominen}, I., \& {Ilyin}, I. 2000,
  \aap, 360, 499

\bibitem[{{Yong} {et~al.}(2005){Yong}, {Carney}, \& {de Almeida}}]{yong05}
{Yong}, D., {Carney}, B.~W., \& {de Almeida}, M.~L.~T. 2005, \aj, 130, 597

\bibitem[{{Zwitter} {et~al.}(2004){Zwitter}, {Castelli}, \& {Munari}}]{zwitter}
{Zwitter}, T., {Castelli}, F., \& {Munari}, U. 2004, \aap, 417, 1055

\end{thebibliography}

\newpage
\clearpage

\begin{figure}
\begin{center}
\includegraphics[scale=0.7, angle=0]{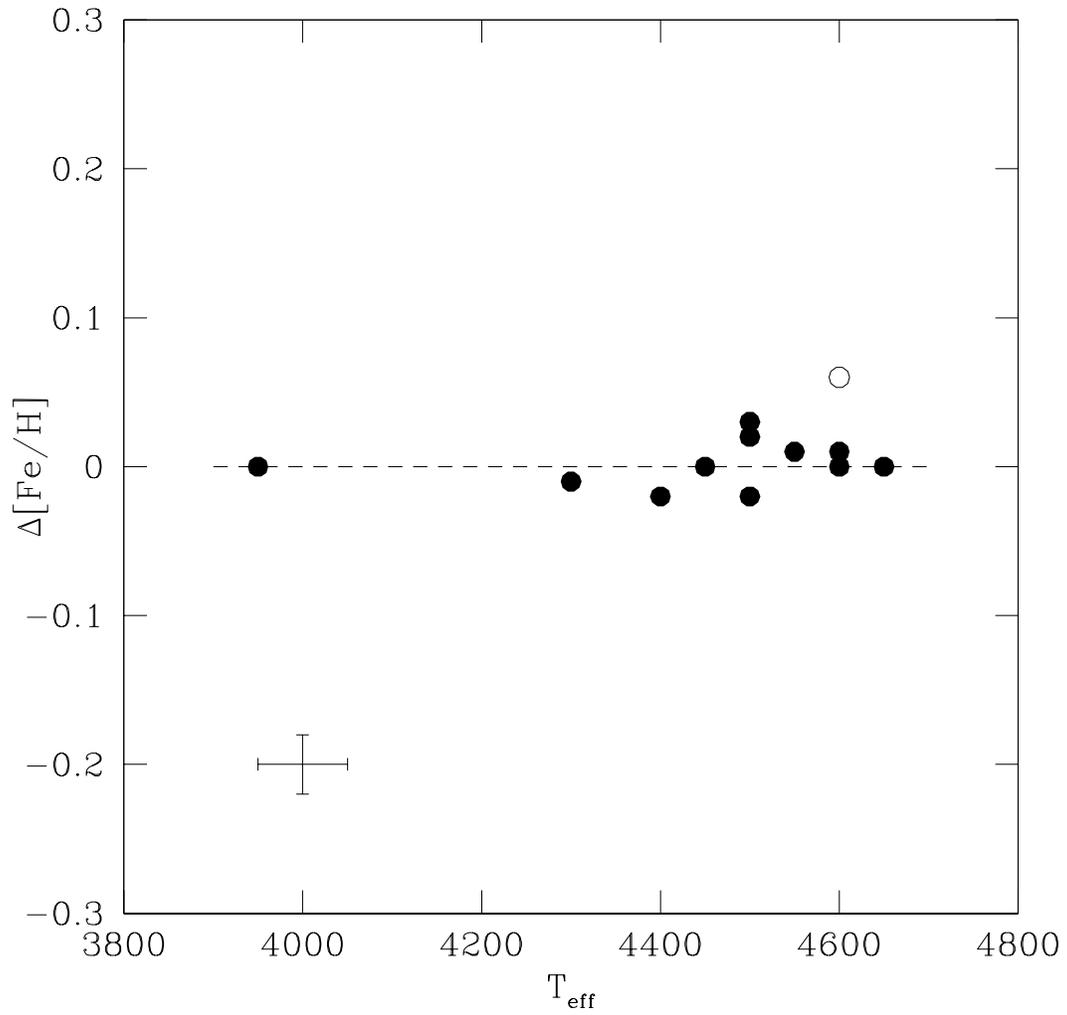}
\caption{Differential Fe abundances vs.~effective temperature for Cr 261. The open circle is the non-member star 2311 as discussed in the text. The dashed line is the cluster mean value. Typical error bars are shown on the bottom left corner.} \label{cr261fe}
\end{center}
\end{figure}

\begin{figure}
\begin{center}
\includegraphics[scale=0.75, angle=0]{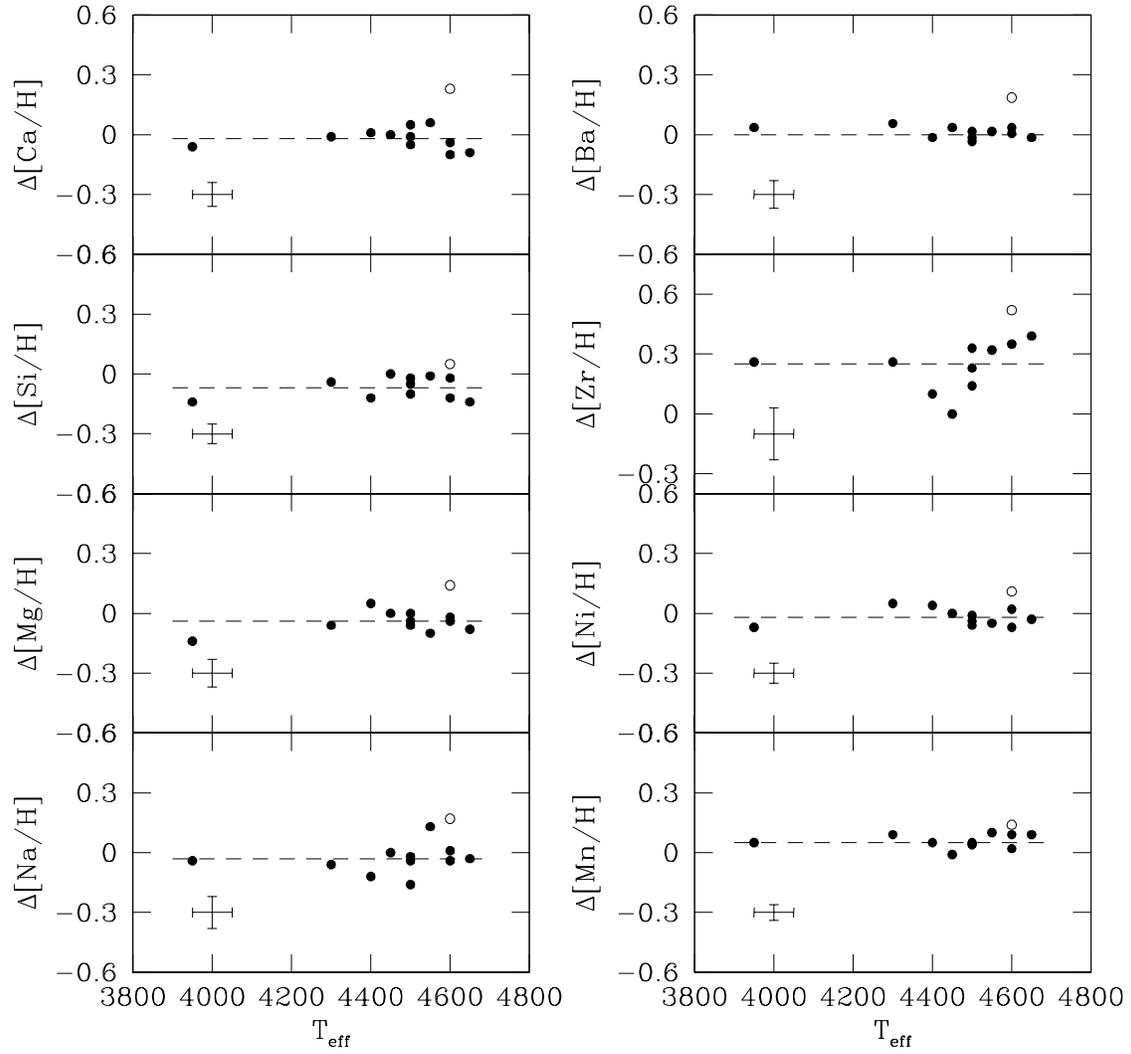}
\caption{Differential [X/H] vs.~effective temperature for various elements. The open circle represents the star 2311. The dashed line is the cluster mean value. Typical error bars are shown at the bottom left corner.} \label{cr261light}
\end{center}
\end{figure}

\begin{figure}
\begin{center}
\includegraphics[scale=0.75, angle=0]{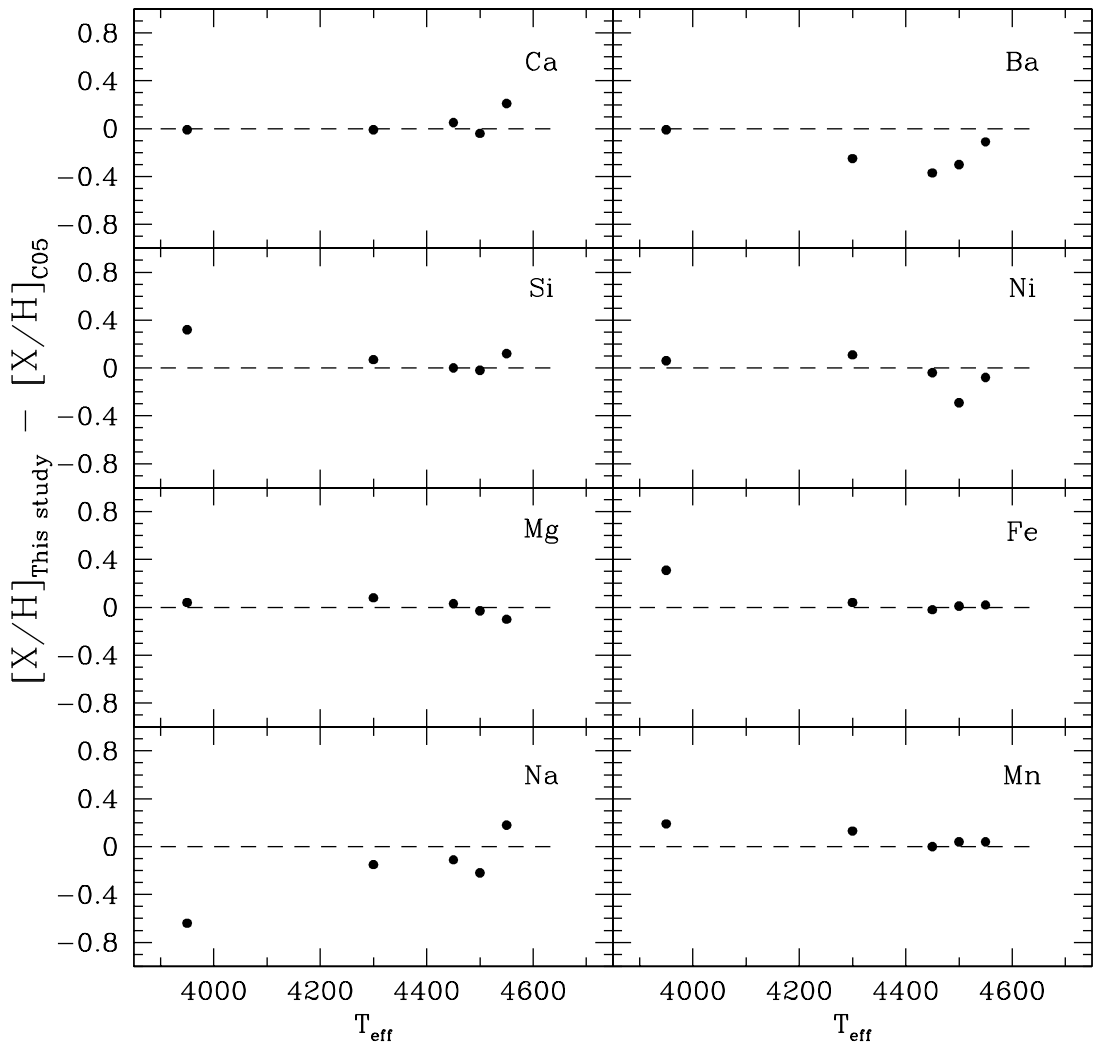}
\caption{A comparison of our results with C05 for the common stars.} \label{cr261comp}
\end{center}
\end{figure}

\begin{figure}
\begin{center}
\includegraphics[scale=0.65]{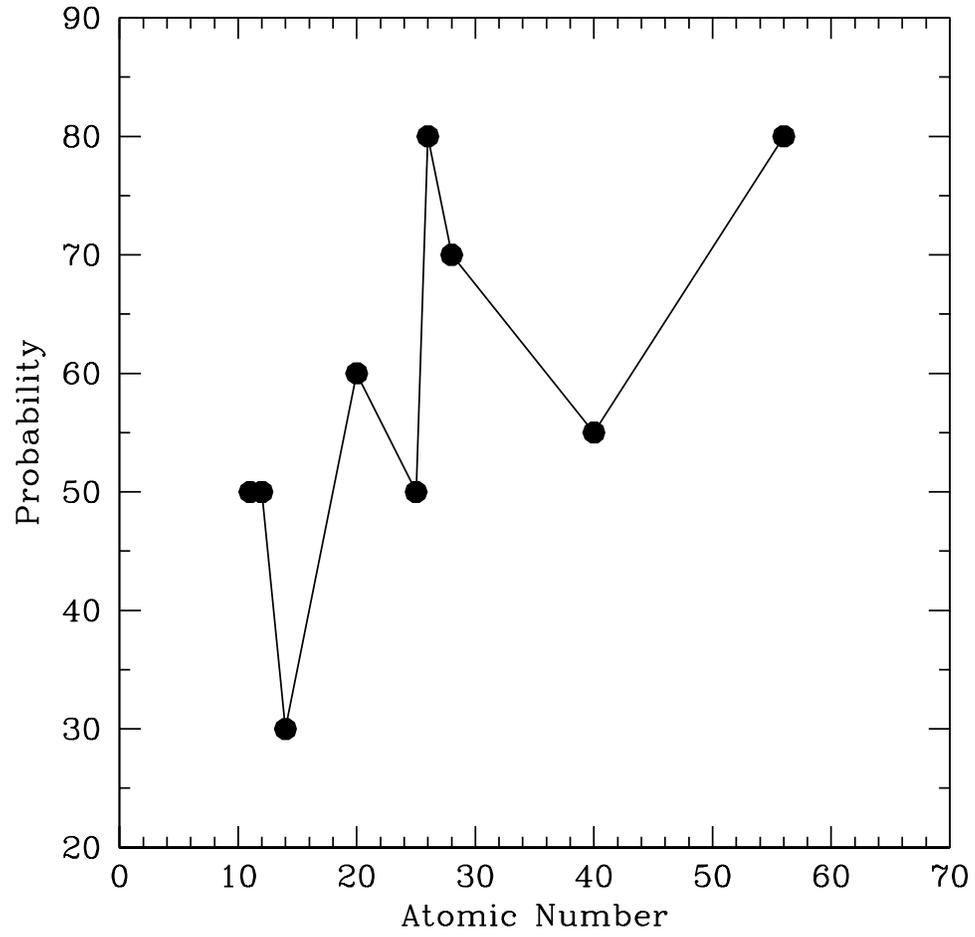}
\caption{The probability of finding a scatter as large as observed given the measuring errors and zero intrinsic scatter for the studied elements. Fe (N = 26) and Ba (N = 56) are the elements most consistent with zero intrinsic scatter.}\label{probcr261}
\end{center}
\end{figure}

\begin{figure}
\begin{center}
\includegraphics[scale=0.75, angle=0]{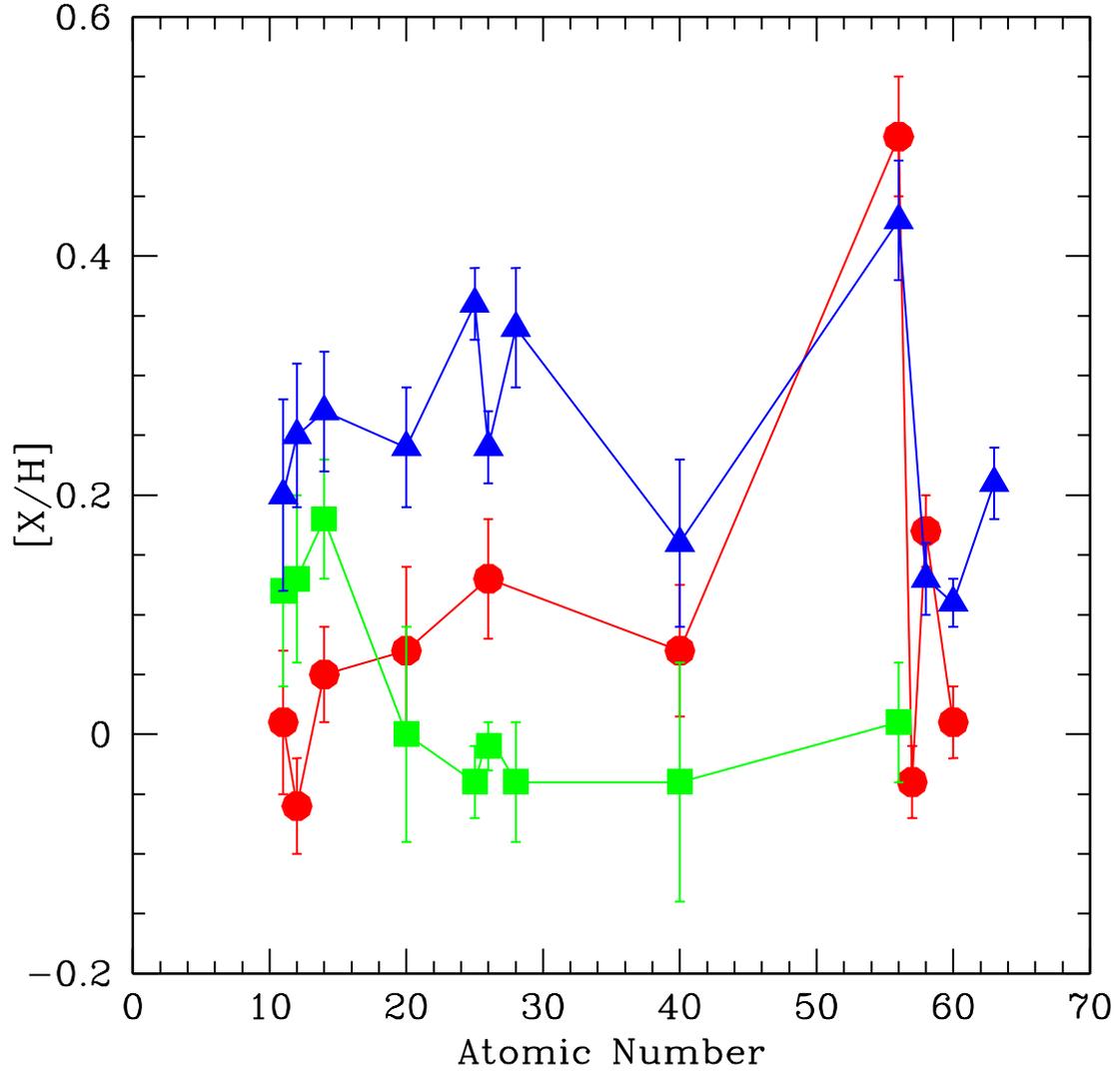}
\caption{Abundance patterns of the three studied clusters. The Hyades is shown in red circles, Collinder 261 is in green squares, and the HR1614 moving group is in blue triangles. The three cluster have their own abundance patterns.}\label{track}
\end{center}
\end{figure}

\begin{figure}
\begin{center}
\includegraphics[scale=0.75, angle=0]{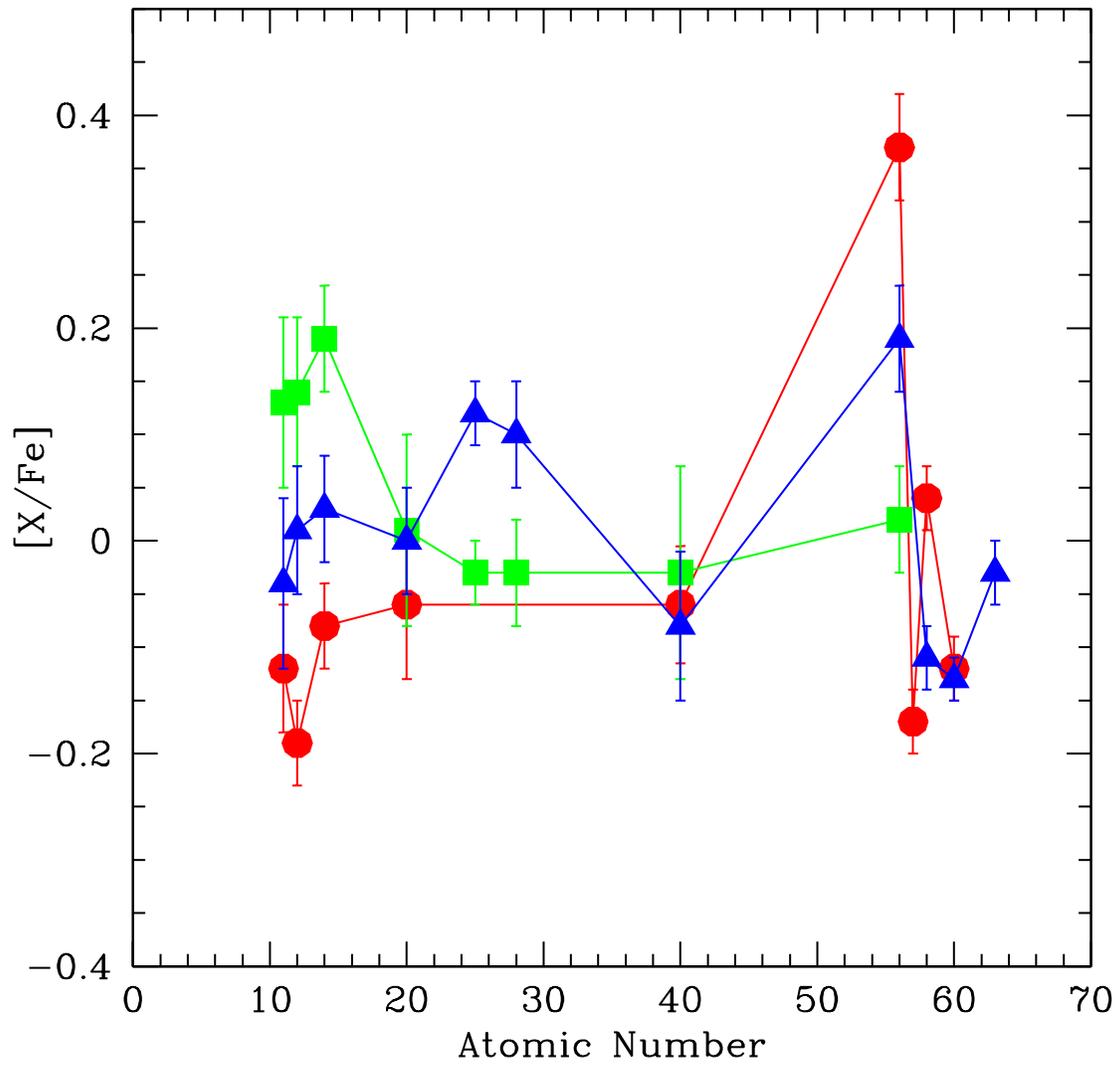}
\caption{Abundances relative to Fe for the three studied clusters. The symbols are same as those in Figure \ref{track}. The three clusters still show unique signatures, demonstrating that abundance patterns are not always locked to Fe. Note Fe is not plotted.}\label{trackfe}
\end{center}
\end{figure}

\begin{figure}
\begin{center}
\includegraphics[scale=0.7, angle=0]{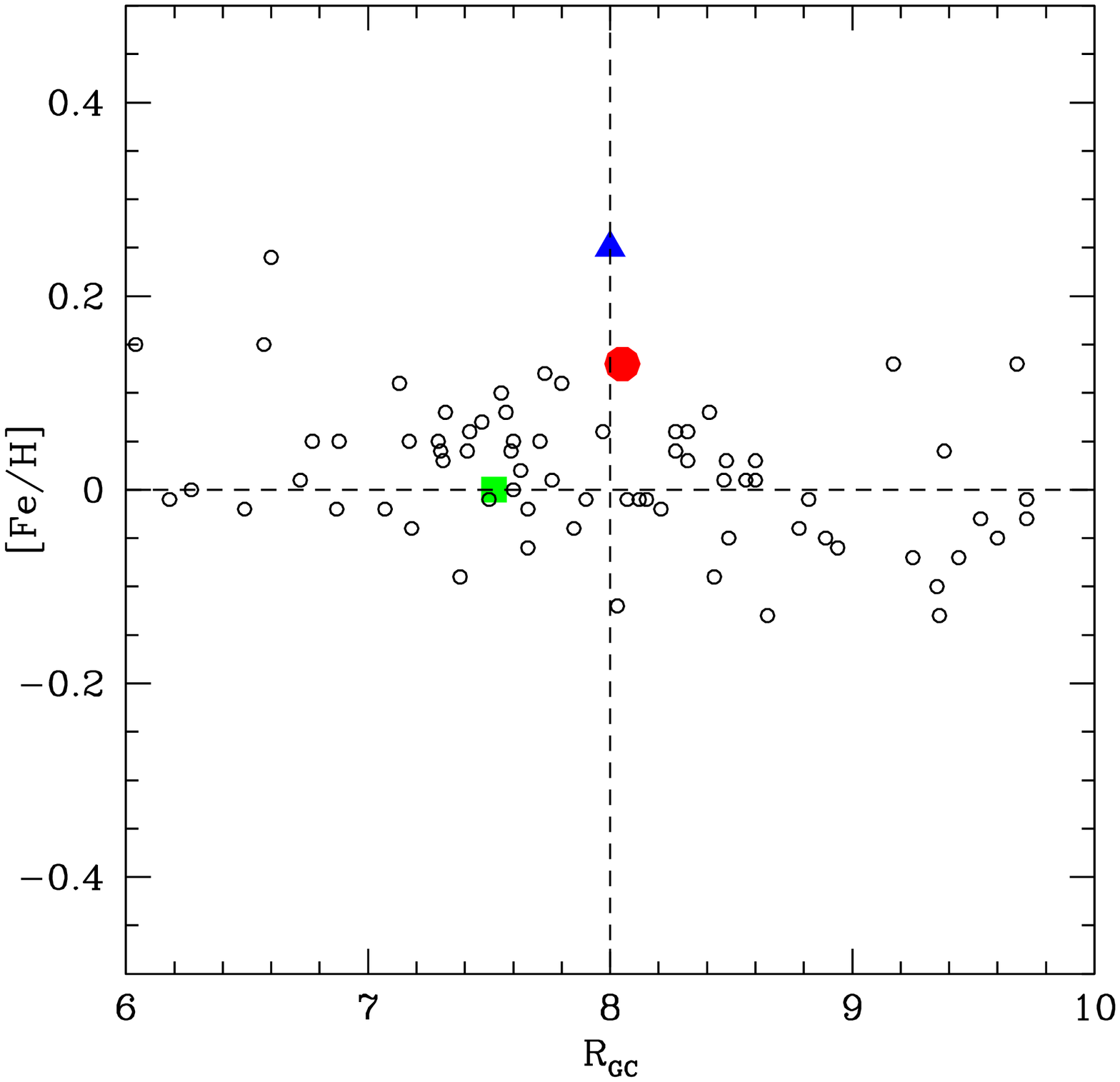}
\caption{Metallicity vs.~Galactocentric radius for cepheids and the studied clusters. The open circles are the cepheid abundances from \citet{cepheids}. The cluster symbols are same as those in Figure \ref{track}. The dashed lines mark the Sun.}\label{ceph_fe}
\end{center}
\end{figure}

\begin{figure}
\begin{center}
\includegraphics[scale=0.8, angle=0]{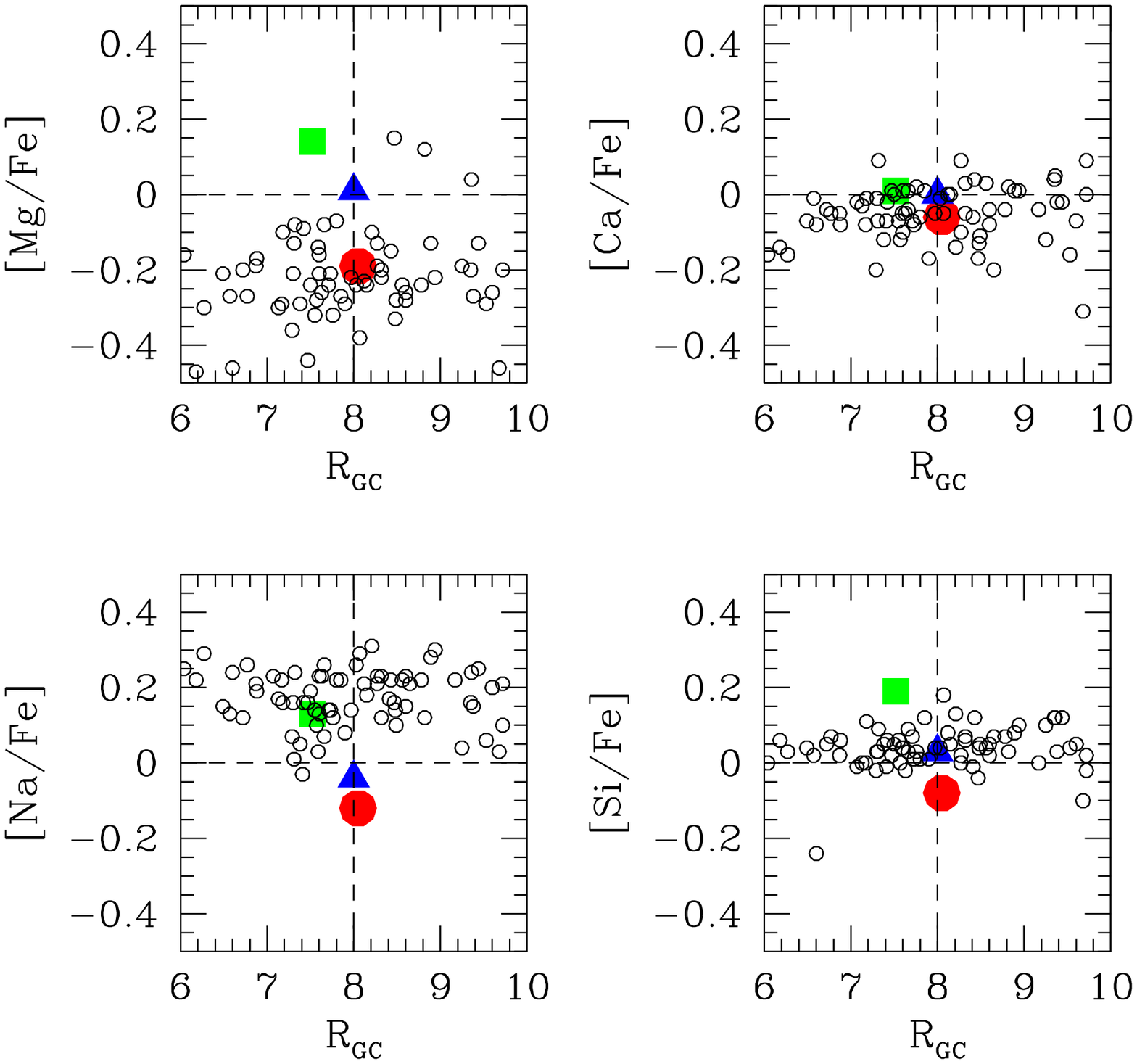}
\caption{[X/Fe] vs.~Galactocentric radius for cepheids and the studied clusters.The open circles are the cepheid abundances from \citet{cepheids}. The cluster symbols are same as those in Figure \ref{track}. The dashed lines mark the Sun.}\label{ceph_alpha}
\end{center}
\end{figure}

\begin{figure}
\begin{center}
\includegraphics[scale=0.8, angle=0]{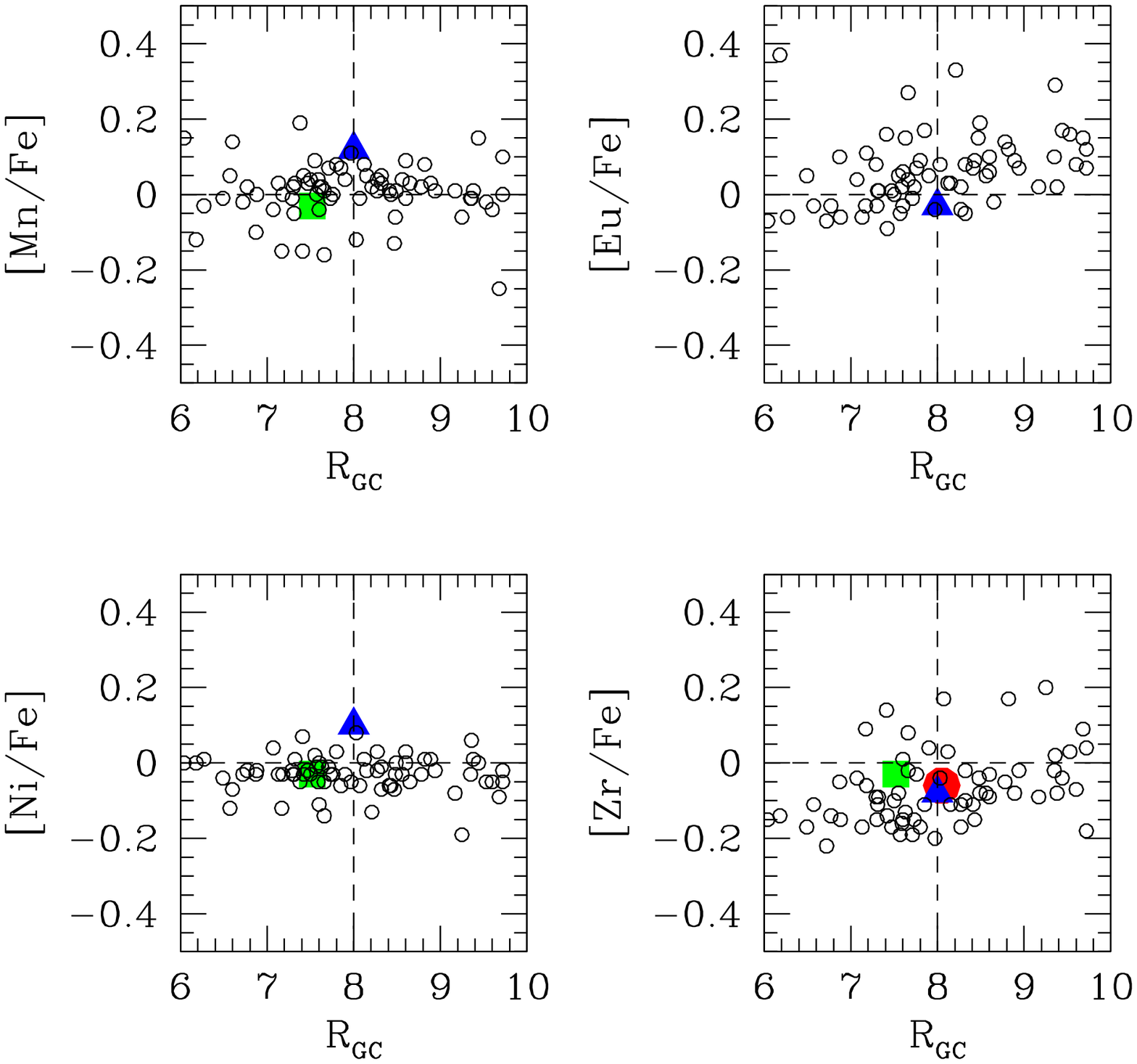}
\caption{[X/Fe] vs.~Galactocentric radius for cepheids and the studied clusters.The open circles are the cepheid abundances from \citet{cepheids}. The cluster symbols are same as those in Figure \ref{track}. The dashed lines mark the Sun.}\label{ceph_heavy}
\end{center}
\end{figure}

\begin{figure}
\begin{center}
\includegraphics[scale=0.75, angle=0]{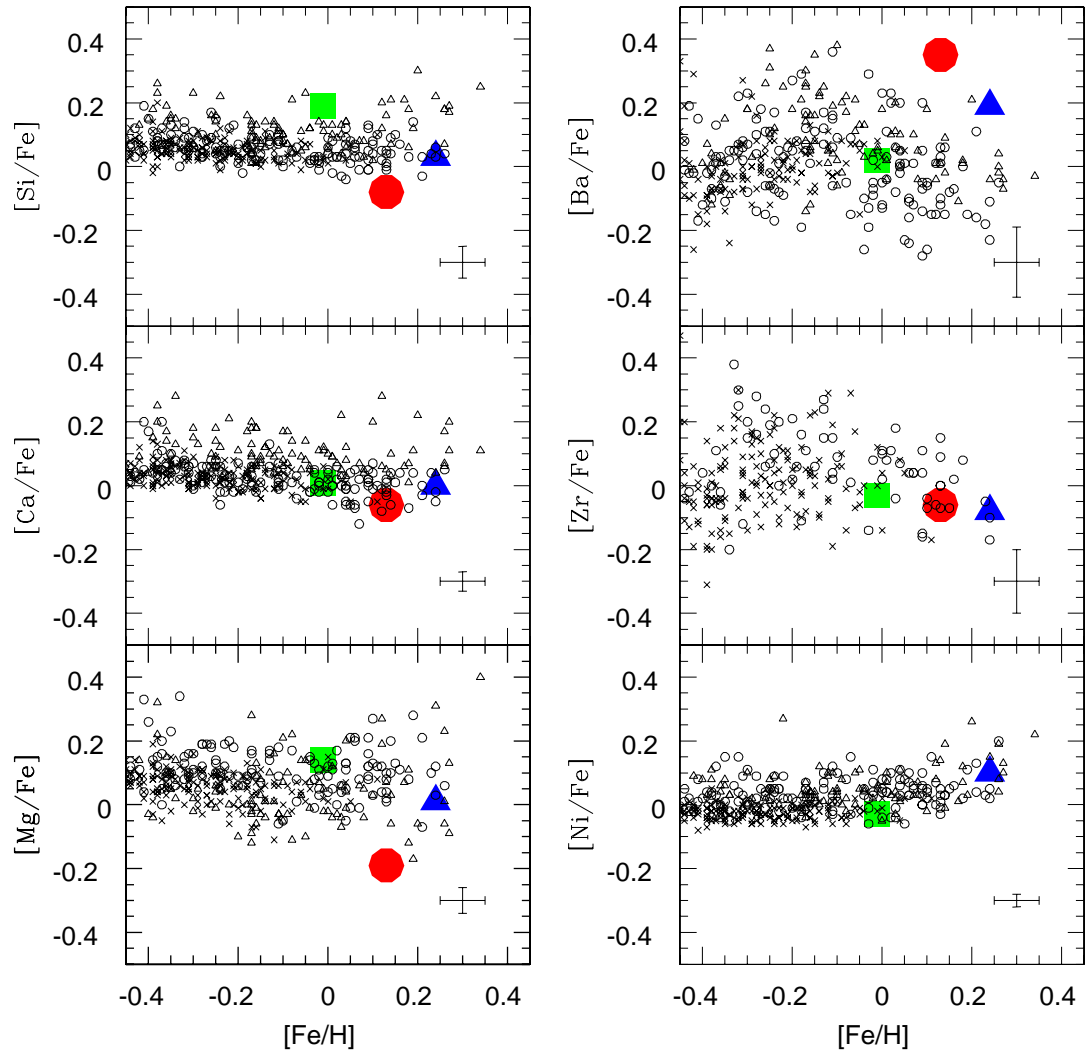}
\caption{Cluster abundances compared to the field surveys. The open circles represent \citet{reddy03} values, crosses represent \citet{ed93} values, and triangles represent \citet{ap04} values. The cluster symbols are same as those in Figure \ref{track}. The error bars are from \citet{reddy03}.}\label{field}
\end{center}
\end{figure}

\begin{figure}
\begin{center}
\includegraphics[scale=0.55, angle=0]{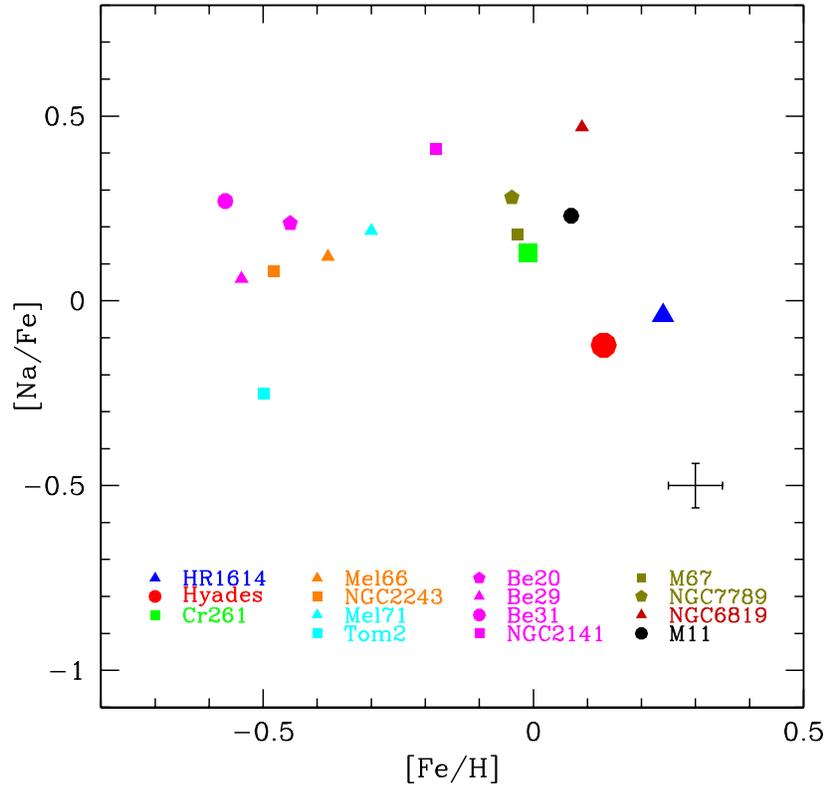}
\includegraphics[scale=0.55, angle=0]{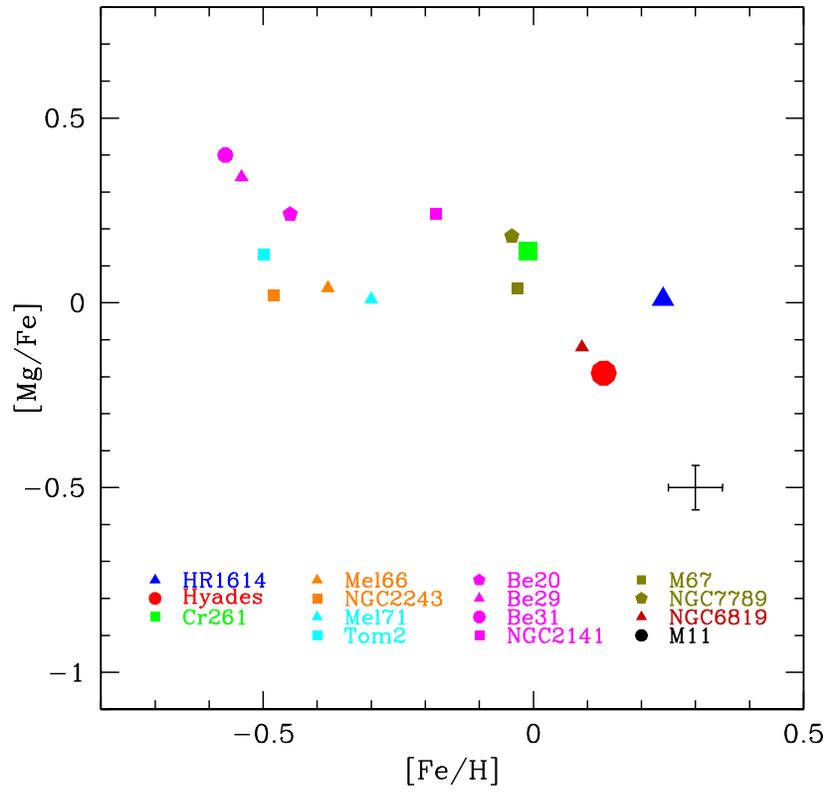}
\caption{Na and Mg abundances for open clusters.}\label{ocna}
\end{center}
\end{figure}

\begin{figure}
\begin{center}
\includegraphics[scale=0.55, angle=0]{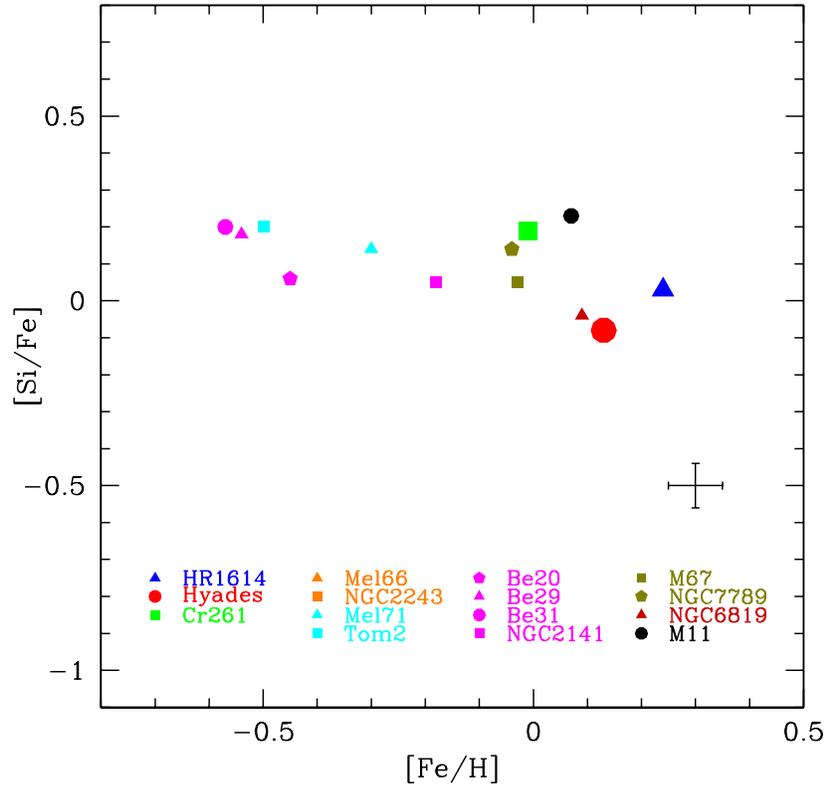}
\includegraphics[scale=0.55, angle=0]{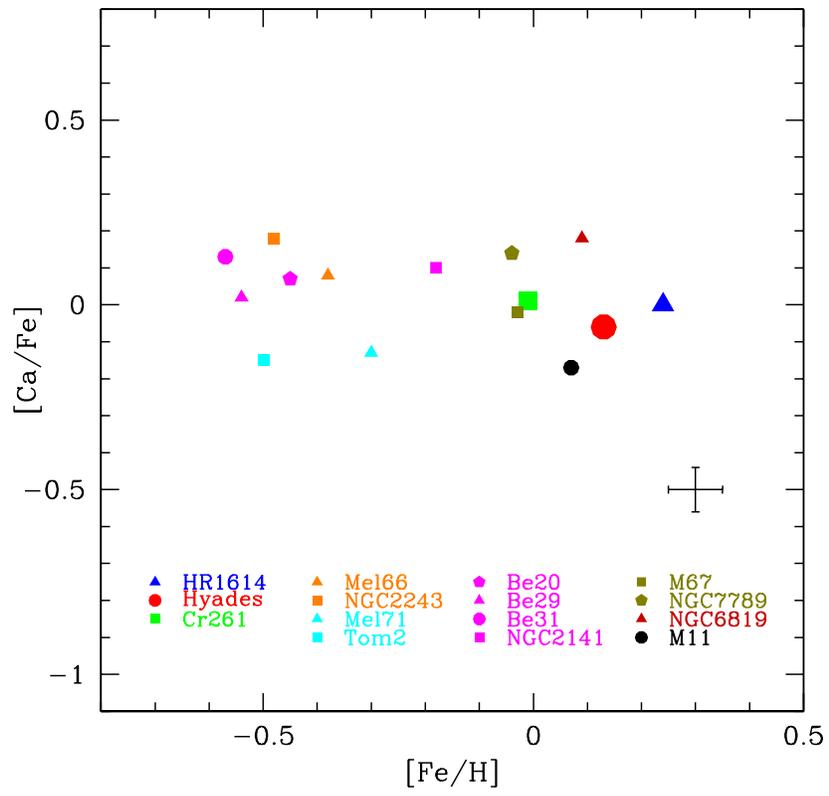}
\caption{Si and Ca abundances for open clusters.}\label{ocsi}
\end{center}
\end{figure}

\begin{figure}
\begin{center}
\includegraphics[scale=0.55, angle=0]{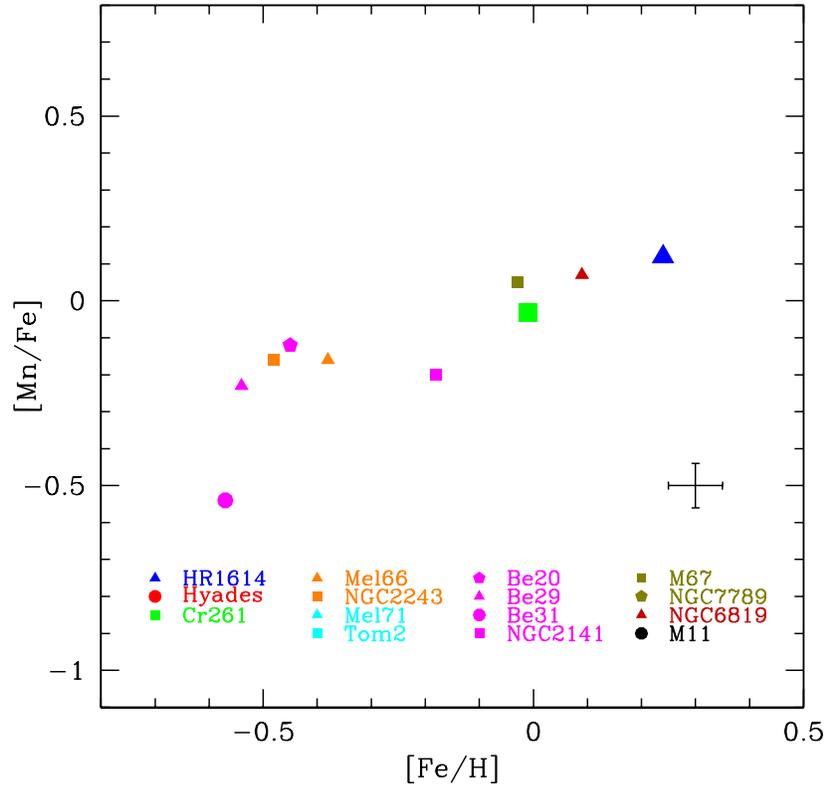}
\includegraphics[scale=0.55, angle=0]{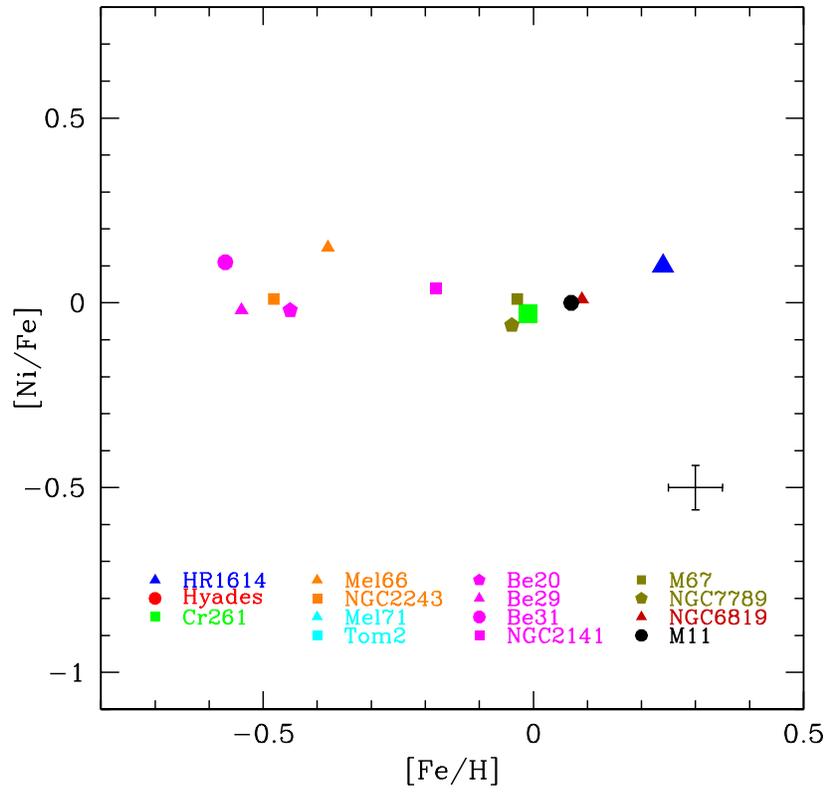}
\caption{Mn and Ni abundances for open clusters.}\label{ocmn}
\end{center}
\end{figure}

\begin{figure}
\begin{center}
\includegraphics[scale=0.55, angle=0]{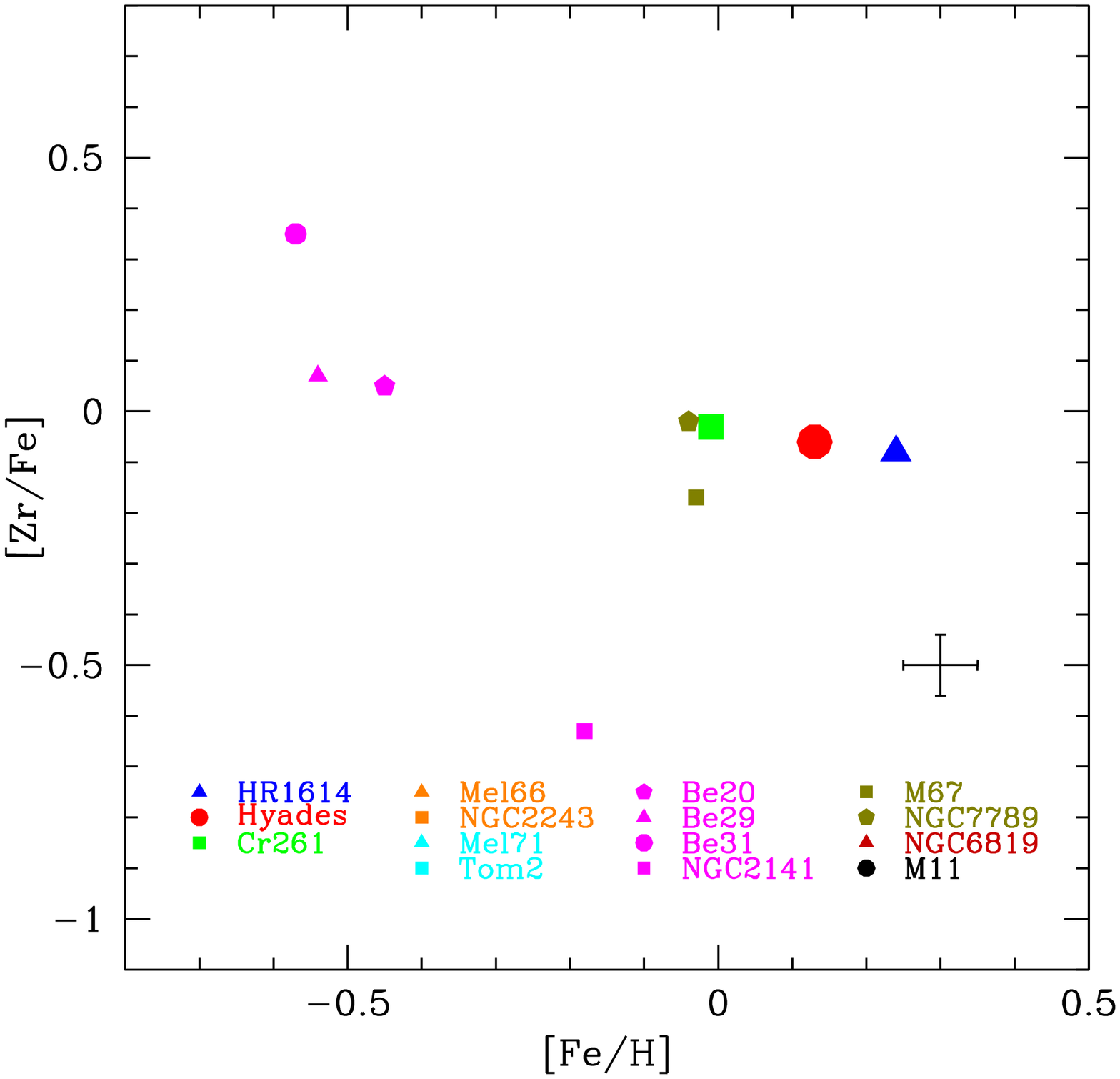}\label{oczr}
\includegraphics[scale=0.55, angle=0]{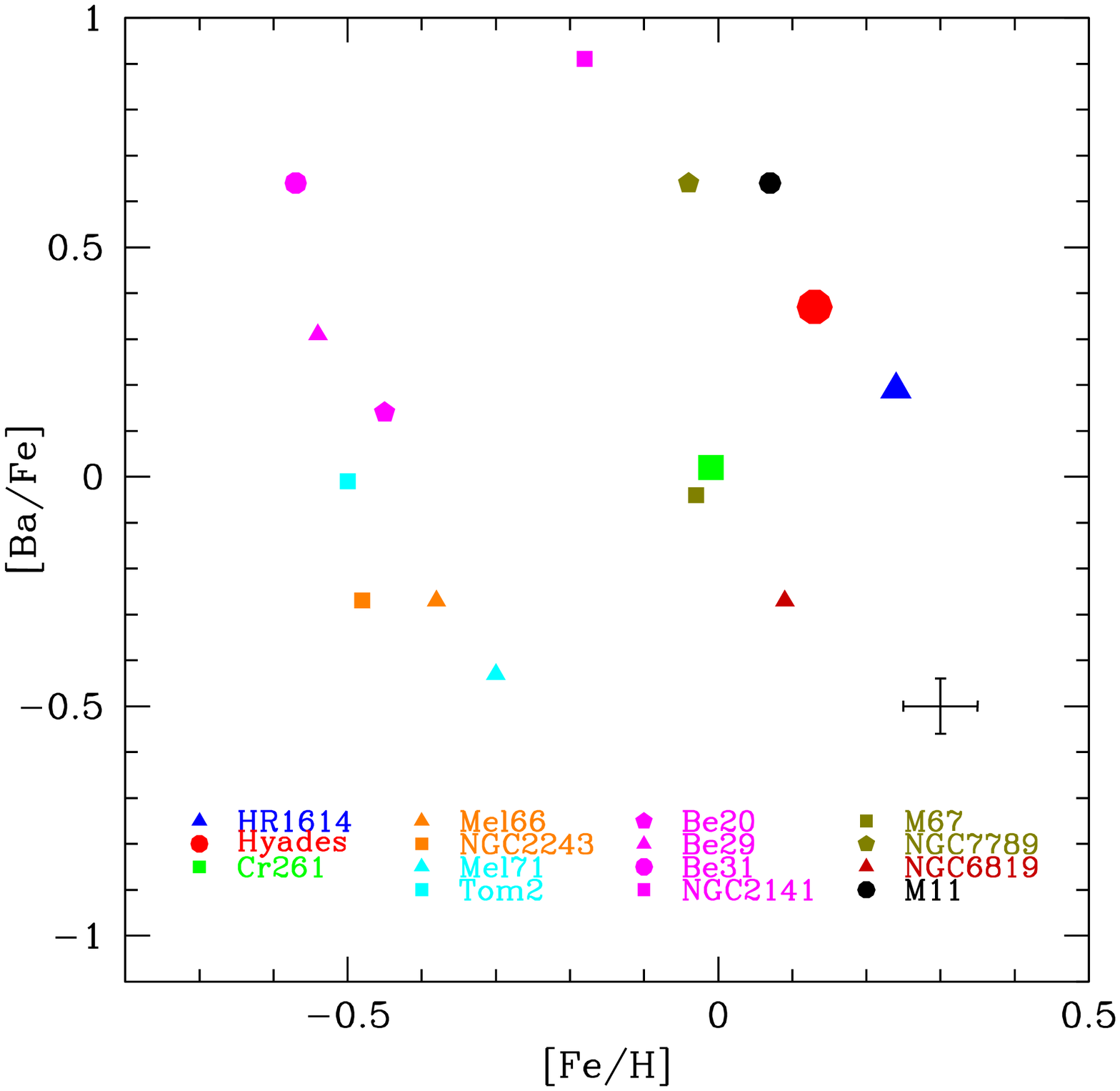}
\caption{The s-process element abundances for open clusters.}\label{ocba}
\end{center}
\end{figure}

\begin{figure}
\begin{center}
\includegraphics[scale=0.55, angle=0]{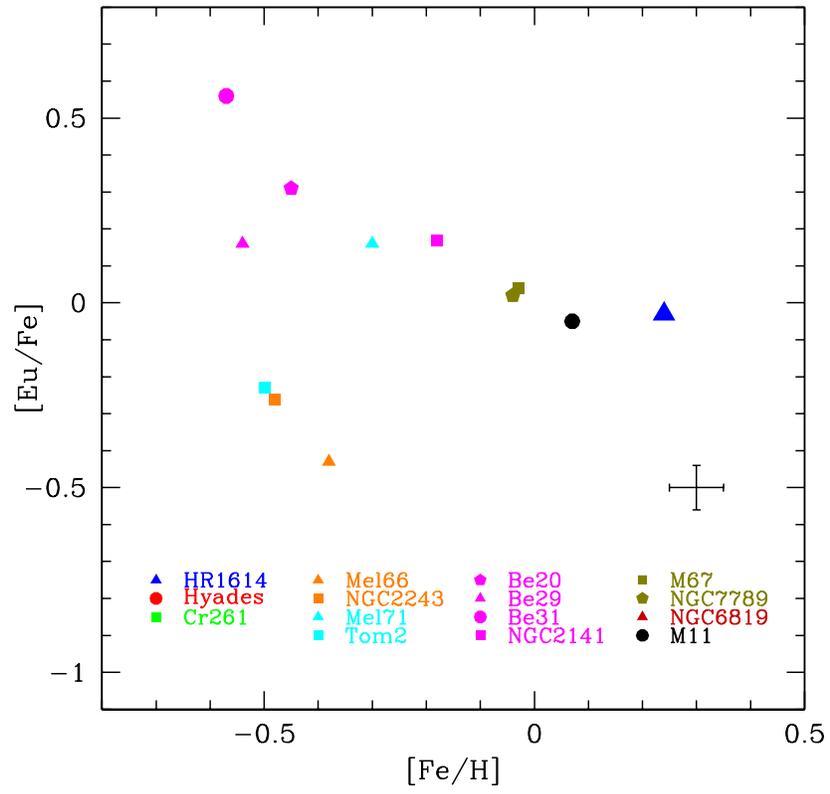}
\caption{Eu (r-process) abundances for open clusters.}\label{oceu}
\end{center}
\end{figure}

\clearpage

\begin{deluxetable}{ccccccccc}
\tabletypesize{\normalsize}
\tablecaption{Collinder 261 Stellar Sample\label{cr261sample}
}
\tablewidth{0pt}
\tablehead{
\colhead{ID (G96)} & 
\colhead{ID (PJM)} & \colhead{RA} & \colhead{DEC} & 
\colhead{V} & \colhead{V-I}  
}

\startdata

2268 & 2001 &  12 37 38.681 & -68 20 25.87 & 13.99 & 1.37     \\
2277 & 1801 &  12 37 45.396 & -68 24  1.02 & 13.61 & 1.36    \\
2285 & 1526 &  12 37 54.105 & -68 21 48.48 & 14.27 & 1.42     \\
2288 & 1481 &  12 37 55.375 & -68 22 35.76 & 13.95 & 1.51    \\
2289 & 1485 &  12 37 55.557 & -68 20 14.36 & 13.74 & 1.60    \\
2291 & 1472 &  12 37 55.504 & -68 24 49.52 & 13.51 & 1.32   \\
2306 & 1080 &  12 38  7.420 & -68 22 30.82 & 13.95 & 1.48    \\
2307 & 1045 &  12 38  8.499 & -68 21 15.01 & 13.58 & 1.53     \\
2311 &  906 &  12 38 12.512 & -68 20 31.45 & 14.21 & 1.44   \\
3027 &   27 &  12 38 40.824 & -68 23 39.13 & 13.89 & 1.66    \\
3029 &   29 &  12 38 40.772 & -68 23 55.46 & 14.31 & 1.49 \\ 
3709 & 1871 &  12 37 43.608 & -68 19 55.06 & 12.43 & 1.84 

\enddata
\tablenotetext{}{Photometry from \citet[PJM]{pjm} and \citet[G96]{gozzoli}}

\end{deluxetable}

\begin{deluxetable}{lccrc|lccrc|lccr} 
\tabletypesize{\scriptsize}
\tablecolumns{14} 
\tablewidth{0pc} 
\tablecaption{Line list\label{tab:line}}
\tablehead{ 
\colhead{Wavelength(\AA)} &
\colhead{Species} &
\colhead{LEP(eV)} &
\colhead{log $gf$} &
\colhead{} &
\colhead{Wavelength(\AA)} &
\colhead{Species} &
\colhead{LEP(eV)} &
\colhead{log $gf$} &
\colhead{} &
\colhead{Wavelength(\AA)} &
\colhead{Species} &
\colhead{LEP(eV)} &
\colhead{log $gf$} 
}
\startdata
5688.19 & Na {\sc i} & 2.11 & $-$0.420 & & 4489.74 & Fe {\sc i} & 0.12 & $-$3.966 & & 6082.72 & Fe {\sc i} & 2.22 & $-$3.650 \\
6154.23 & Na {\sc i} & 2.10 & $-$1.530 & & 4494.57 & Fe {\sc i} & 2.20 & $-$1.136 & & 6093.64 & Fe {\sc i} & 4.61 & $-$1.510 \\
6160.75 & Na {\sc i} & 2.10 & $-$1.230 & & 4523.40 & Fe {\sc i} & 3.65 & $-$1.990 & & 6094.37 & Fe {\sc i} & 4.65 & $-$1.650 \\
4571.09 & Mg {\sc i} & 0.00 & $-$5.393 & & 4531.15 & Fe {\sc i} & 1.48 & $-$2.155 & & 6096.66 & Fe {\sc i} & 3.98 & $-$1.880 \\
4702.99 & Mg {\sc i} & 4.33 & $-$0.380 & & 4531.58 & Fe {\sc i} & 3.93 & $-$2.059 & & 6105.13 & Fe {\sc i} & 4.55 & $-$1.990 \\
5711.09 & Mg {\sc i} & 4.35 & $-$1.833 & & 4547.85 & Fe {\sc i} & 3.54 & $-$1.012 & & 6120.24 & Fe {\sc i} & 0.91 & $-$5.970 \\
5665.56 & Si {\sc i} & 4.92 & $-$1.940 & & 4556.93 & Fe {\sc i} & 3.25 & $-$2.710 & & 6151.62 & Fe {\sc i} & 2.17 & $-$3.299 \\
5684.49 & Si {\sc i} & 4.95 & $-$1.550 & & 4561.43 & Fe {\sc i} & 2.76 & $-$3.080 & & 6157.73 & Fe {\sc i} & 4.08 & $-$1.320 \\
5690.43 & Si {\sc i} & 4.93 & $-$1.770 & & 4593.52 & Fe {\sc i} & 3.94 & $-$2.060 & & 6159.38 & Fe {\sc i} & 4.61 & $-$1.970 \\
5948.54 & Si {\sc i} & 5.08 & $-$1.230 & & 4602.01 & Fe {\sc i} & 1.61 & $-$3.154 & & 6173.34 & Fe {\sc i} & 2.22 & $-$2.880 \\
6142.48 & Si {\sc i} & 5.62 & $-$1.540 & & 5379.57 & Fe {\sc i} & 3.68 & $-$1.730 & & 6180.20 & Fe {\sc i} & 2.73 & $-$2.637 \\
6145.01 & Si {\sc i} & 5.62 & $-$1.362 & & 5417.03 & Fe {\sc i} & 4.41 & $-$1.550 & & 6200.31 & Fe {\sc i} & 2.61 & $-$2.437 \\
6155.13 & Si {\sc i} & 5.62 & $-$0.786 & & 5466.99 & Fe {\sc i} & 3.57 & $-$2.440 & & 4491.41 & Fe {\sc ii}& 2.86 & $-$2.684 \\
4455.88 & Ca {\sc i} & 1.89 & $-$0.526 & & 5618.63 & Fe {\sc i} & 4.21 & $-$1.292 & & 4508.28 & Fe {\sc ii}& 2.86 & $-$2.312 \\
4578.55 & Ca {\sc i} & 2.25 & $-$0.558 & & 5633.95 & Fe {\sc i} & 4.99 & $-$0.270 & & 4541.52 & Fe {\sc ii}& 2.84 & $-$2.990 \\
5867.57 & Ca {\sc i} & 2.93 & $-$1.570 & & 5662.52 & Fe {\sc i} & 4.16 & $-$0.520 & & 4576.33 & Fe {\sc ii}& 2.84 & $-$2.822 \\
6169.04 & Ca {\sc i} & 2.52 & $-$0.797 & & 5701.55 & Fe {\sc i} & 2.56 & $-$2.216 & & 4582.83 & Fe {\sc ii}& 2.84 & $-$3.094 \\
6169.56 & Ca {\sc i} & 2.53 & $-$0.478 & & 5705.47 & Fe {\sc i} & 4.30 & $-$1.420 & & 4620.52 & Fe {\sc ii}& 2.83 & $-$3.079 \\
6013.53 & Mn {\sc i} & 3.07 & $-$0.251 & & 5741.85 & Fe {\sc i} & 4.25 & $-$1.689 & & 4656.98 & Fe {\sc ii}& 2.89 & $-$3.552 \\
6016.67 & Mn {\sc i} & 3.08 & $-$0.100 & & 5775.08 & Fe {\sc i} & 4.22 & $-$1.310 & & 4670.17 & Fe {\sc ii}& 2.58 & $-$3.904 \\
6021.80 & Mn {\sc i} & 3.08 &    0.034 & & 5778.45 & Fe {\sc i} & 2.59 & $-$3.480 & & 5414.08 & Fe {\sc ii}& 3.22 & $-$3.680 \\
4216.19 & Fe {\sc i} & 0.00 & $-$3.356 & & 5811.92 & Fe {\sc i} & 4.14 & $-$2.430 & & 5991.38 & Fe {\sc ii}& 3.15 & $-$3.557 \\
4222.22 & Fe {\sc i} & 2.45 & $-$0.967 & & 5837.70 & Fe {\sc i} & 4.29 & $-$2.340 & & 6084.11 & Fe {\sc ii}& 3.20 & $-$3.808 \\
4232.72 & Fe {\sc i} & 0.11 & $-$4.928 & & 5853.16 & Fe {\sc i} & 1.49 & $-$5.280 & & 6149.26 & Fe {\sc ii}& 3.89 & $-$2.724 \\
4233.61 & Fe {\sc i} & 2.48 & $-$0.604 & & 5855.09 & Fe {\sc i} & 4.60 & $-$1.547 & & 5846.99 & Ni {\sc i} & 1.68 & $-$3.210 \\
4237.09 & Fe {\sc i} & 0.95 & $-$4.379 & & 5856.10 & Fe {\sc i} & 4.29 & $-$1.640 & & 6086.28 & Ni {\sc i} & 4.26 & $-$0.515 \\
4347.24 & Fe {\sc i} & 0.00 & $-$5.503 & & 5858.79 & Fe {\sc i} & 4.22 & $-$2.260 & & 6175.37 & Ni {\sc i} & 4.09 & $-$0.535 \\
4375.93 & Fe {\sc i} & 0.00 & $-$3.031 & & 5927.80 & Fe {\sc i} & 4.65 & $-$1.090 & & 6177.24 & Ni {\sc i} & 1.83 & $-$3.510 \\
4389.24 & Fe {\sc i} & 0.05 & $-$4.583 & & 5956.69 & Fe {\sc i} & 0.86 & $-$4.608 & & 6127.44 & Zr {\sc i} & 0.15 & $-$1.060 \\
4439.88 & Fe {\sc i} & 2.28 & $-$3.002 & & 6015.24 & Fe {\sc i} & 2.22 & $-$4.760 & & 6134.55 & Zr {\sc i} & 0.00 & $-$1.280 \\
4442.34 & Fe {\sc i} & 2.19 & $-$1.255 & & 6024.06 & Fe {\sc i} & 4.54 &    0.610 & & 6143.20 & Zr {\sc i} & 0.07 & $-$1.100 \\
4445.48 & Fe {\sc i} & 0.08 & $-$5.441 & & 6034.03 & Fe {\sc i} & 4.31 & $-$2.470 & & 5853.69 & Ba {\sc ii}& 0.60 & $-$1.010 \\
4447.72 & Fe {\sc i} & 2.22 & $-$1.342 & & 6042.22 & Fe {\sc i} & 4.65 & $-$0.890 & & 6141.73 & Ba {\sc ii}& 0.70 & $-$0.070 \\
4427.31 & Fe {\sc i} & 0.05 & $-$3.044 & & 6054.08 & Fe {\sc i} & 4.37 & $-$2.310 & & \nodata & \nodata & \nodata & \nodata \\
4461.65 & Fe {\sc i} & 0.08 & $-$5.441 & & 6056.00 & Fe {\sc i} & 4.73 & $-$0.650 & & \nodata & \nodata & \nodata & \nodata
\enddata
\end{deluxetable}

\begin{deluxetable}{lccccccccccc}
\tabletypesize{\normalsize}
\tablecaption{Stellar Parameters \label{cr261params}
}
\tablewidth{0pt}
\tablehead{
\colhead{ID}& 
\colhead{T$_{eff}$} & \colhead{Log $g$} & \colhead{$\xi$} &
\colhead{T$_{eff}$ (C05)} & \colhead{Log $g$ (C05)} & \colhead{$\xi$(C05)} &
\colhead{T$_{eff}$ (F03)} & \colhead{Log $g$ (F03)} & \colhead{$\xi$(F03)}
}

\startdata
2268 & 4550 & 2.0 & 1.52  & 4580    & 1.83  & 1.26   &\nodata&\nodata&\nodata\\
2277 & 4600 & 2.0  & 1.4  & \nodata &\nodata&\nodata &\nodata&\nodata&\nodata\\
2285 & 4600 & 2.0  & 0.9  & \nodata &\nodata&\nodata &\nodata&\nodata&\nodata\\
2288 & 4400 & 2.1  & 1.2  & \nodata &\nodata&\nodata &\nodata&\nodata&\nodata\\
2289 & 4300 & 1.8  & 1.4  & 4340    & 1.76  & 1.27   &\nodata&\nodata&\nodata\\
2291 & 4650 & 2.3  & 1.8  & \nodata &\nodata&\nodata &\nodata&\nodata&\nodata\\
2306 & 4500 & 2.1  & 1.5  & 4500    & 2.09  & 1.23   & 4900 & 2.2 & 1.2 \\
2307 & 4450 & 1.8  & 1.1  & 4470    & 2.07  & 1.23   & 4400 & 1.5 & 1.2 \\
2311 & 4600 & 2.0  & 0.9  & \nodata &\nodata&\nodata &\nodata&\nodata&\nodata\\
3027 & 4500 & 2.0  & 1.2  & \nodata &\nodata&\nodata &\nodata&\nodata&\nodata\\
3029 & 4500 & 1.9  & 1.4  & \nodata &\nodata&\nodata &\nodata&\nodata&\nodata\\
3709 & 3950 & 0.5  & 1.4  & 3980    & 0.43  & 1.44   & 4000 & 0.7 & 1.5
\enddata

\end{deluxetable}

\begin{deluxetable}{lccccccccccccccccc}
\tabletypesize{\normalsize}
\tablecaption{Absolute Abundances (log $\epsilon$)\label{ab_table}
}
\tablewidth{0pt}
\tablehead{
\colhead{ID}& 
\colhead{Na} & \colhead{Mg} & \colhead{Si} & 
\colhead{Ca} & \colhead{Mn} & \colhead{Fe}
& \colhead{Ni} & \colhead{Zr} & \colhead{Ba (with HFS)} & \colhead {Ba (without HFS)}
}

\startdata
2268 & 6.61 & 7.65 & 7.79 & 6.44 & 5.40 & 7.52 & 6.17 & 2.63 & 2.15 & 2.36 \\
2277 & 6.49 & 7.78 & 7.78 & 6.28 & 5.39 & 7.51 & 6.15 & 2.66 & 2.17 & 2.27 \\
2285 & 6.44 & 7.71 & 7.68 & 6.34 & 5.32 & 7.52 & 6.24 & 2.66 & 2.14 & 2.34 \\
2288 & 6.36 & 7.80 & 7.68 & 6.39 & 5.35 & 7.49 & 6.26 & 2.41 & 2.13 & 2.34 \\
2289 & 6.42 & 7.69 & 7.76 & 6.37 & 5.39 & 7.50 & 6.27 & 2.57 & 2.17 & 2.29 \\
2291 & 6.45 & 7.67 & 7.66 & 6.29 & 5.39 & 7.51 & 6.19 & 2.70 & 2.13 & 2.25 \\
2306 & 6.32 & 7.71 & 7.75 & 6.37 & 5.34 & 7.54 & 6.16 & 2.45 & 2.14 & 2.30 \\
2307 & 6.48 & 7.75 & 7.80 & 6.38 & 5.29 & 7.51 & 6.22 & 2.31 & 2.15 & 2.31 \\
2311 & 6.65 & 7.89 & 7.85 & 6.61 & 5.44 & 7.56 & 6.33 & 2.73 & 2.37 & 2.47 \\
3027 & 6.46 & 7.75 & 7.78 & 6.43 & 5.34 & 7.53 & 6.21 & 2.64 & 2.13 & 2.23 \\
3029 & 6.44 & 7.69 & 7.70 & 6.33 & 5.35 & 7.49 & 6.18 & 2.54 & 2.14 & 2.38 \\
3709 & 6.44 & 7.61 & 7.66 & 6.32 & 5.35 & 7.51 & 6.15 & 2.57 & 2.15 & 2.30 \\
\enddata

\end{deluxetable}

\begin{deluxetable}{crrrrrrrrrrrr}
\tabletypesize{\normalsize}
\tablecaption{Abundance dependencies on model atmospheres (Kurucz - MARCS) \label{models1}
}
\tablewidth{0pt}
\tablehead{\colhead{Star}&
\colhead{$\delta$[Na/H]} & \colhead{$\delta$[Mg/H]} & \colhead{$\delta$[Si/H]} & 
\colhead{$\delta$[Ca/H]} & \colhead{$\delta$[Mn/H]} & \colhead{$\delta$[Fe/H]}
& \colhead{$\delta$[Ni/H]} & \colhead{$\delta$[Zr/H]} & \colhead{$\delta$[Ba/H]}
}

\startdata
2285 (T$_{eff}$ = 4600K)& $-$0.01 & 0.00 & 0.01 & $-$0.01 & $-$ 0.01 & 0.01 & 0.00 & $-$ 0.02 & 0.03 \\
2288 (T$_{eff}$ = 4400K)& 0.04 & $-$ 0.01 & $-$ 0.07 & 0.05 & 0.00 & $-$ 0.03 & $-$ 0.11 & 0.00 & $-$ 0.06 \\
3709 (T$_{eff}$ = 3950K)& 0.15 & 0.01 & 0.02 & 0.17 & 0.05 & 0.01 & 0.02 & 0.35 & 0.06 \\
\enddata
\end{deluxetable}

\begin{deluxetable}{crrrrrrrrrrrrr}
\tabletypesize{\normalsize}
\tablecaption{Abundance dependencies on model atmospheres (Kurucz - MARCS) with adjusted microturbulence \label{models2}
}
\tablewidth{0pt}
\tablehead{\colhead{Star}&\colhead{$\delta$[Na/H]} & \colhead{$\delta$[Mg/H]} & \colhead{$\delta$[Si/H]} & 
\colhead{$\delta$[Ca/H]} & \colhead{$\delta$[Mn/H]} & \colhead{$\delta$[Fe/H]}
& \colhead{$\delta$[Ni/H]} & \colhead{$\delta$[Zr/H]} & \colhead{$\delta$[Ba/H]}
}

\startdata
2285 (T$_{eff}$ = 4600K)& 0.01 & 0.00 & $-$ 0.01 & 0.01 & $-$ 0.02 & $-$ 0.02 & $-$ 0.02 & $-$ 0.02 & 0.00 \\
2288 (T$_{eff}$ = 4400K)& 0.04 & $-$ 0.01 & $-$ 0.04 & 0.05 & 0.00 & $-$ 0.03 & $-$ 0.04 & 0.00 & $-$ 0.06 \\
3709 (T$_{eff}$ = 3950K)& 0.03 & 0.01 & 0.02 & 0.01 & 0.00 & 0.01 & 0.02 & 0.07 & 0.06 \\
\enddata
\end{deluxetable}

\begin{deluxetable}{lccccccccccccccccccccccccc}
\tabletypesize{\normalsize}
\tablecaption{Abundance Sensitivities for star 2268 \label{tab:error}
}
\tablewidth{0pt}
\tablehead{
\colhead{}&\colhead{$\delta$[Fe/H]} &
\colhead{$\delta$[Na/H]} & \colhead{$\delta$[Mg/H]} &
\colhead{$\delta$[Si/H]} &  \colhead{$\delta$[Ca/H]} &
\colhead{$\delta$[Mn/H]} & \colhead{$\delta$[Ni/H]} &
\colhead{$\delta$[Zr/H]} & \colhead{$\delta$[Ba/H]} 
}

\startdata

$\Delta$ T$_{\rm eff} = \pm$50 & $\pm$0.01 & $\pm$0.04 & $\pm$0.02 & $\mp$0.03 & $\pm$0.04 & $\pm$0.00 & $\pm$0.01 & $\pm$0.10 & 0.00 \\
$\Delta$ log $g$ = $\pm$0.1&$\pm$0.02 & $\mp$0.02 & 0.00 & $\pm$0.02 & $\mp$0.00 & $\mp$0.01 & $\pm$0.03 & $\pm$0.01 & $\pm$0.05 \\
$\Delta  \xi = \pm$0.2 &$\mp$0.02 & $\mp$0.04 & $\mp$0.04 & $\mp$0.03 & $\mp$0.05 & $\mp$0.02 & $\mp$0.04& $\mp$0.04 & $\mp$0.05 \\ 
$\Delta$ EW& $\pm$0.01 & $\pm$0.05& $\pm$0.04 & $\pm$0.03 & $\pm$0.02 & $\pm$0.02 & $\pm$0.01 & $\pm$0.07 & $\pm$0.02 \\ 
\hline
Total & $\pm$0.03 &  $\pm$0.08 & $\pm$0.06 & $\pm$0.05 & $\pm$0.06 & $\pm$0.04 & $\pm$0.05 & $\pm$0.13 & $\pm$0.08 \\
\enddata
\end{deluxetable}

\begin{deluxetable}{lcccccc}
\tabletypesize{\normalsize}
\tablecaption{Radial Velocities\label{cr261rv}
}
\tablewidth{0pt}
\tablehead{
\colhead{Star ID} & \colhead{this study} & \colhead{\citet{friel02}} 
}

\startdata
2268 & -24.5& -35  \\
2277 & -24.2& -35  \\
2285 & -26.9& -28  \\
2288 & -27.3& -27  \\
2289 & -27.2& -35  \\
2291 & -27.8& -31  \\
2306 & -28.1& -31  \\
2307 & -26.3& -30  \\
2311 & -18.1& -30  \\
3027 & -24.5& -31  \\
3029 & -24.2& -16  \\
3709 & -28.2& -37  \\

\enddata

\end{deluxetable}

\begin{deluxetable}{lcccccccc}
\tabletypesize{\normalsize}
\tablecaption{Abundance scatter\label{cr261scatter}
}
\tablewidth{0pt}
\tablehead{
\colhead{Element} & \colhead{$\sigma_{obs}$}  & \colhead{$\sigma_{int}$ (upper)} 
}

\startdata

Fe & 0.02 & 0.02 \\
Na & 0.07 & 0.04 \\
Mg & 0.05 & 0.04  \\
Si & 0.06 & 0.05 \\
Ca & 0.05 & 0.04 \\
Mn & 0.03 & 0.02  \\
Ni & 0.04 & 0.03 \\
Zr & 0.12 & 0.05 \\
Ba & 0.03 & 0.02 \\

\enddata

\end{deluxetable}

\begin{deluxetable}{cccccccccc}
\tabletypesize{\normalsize}
\tablecaption{Cluster Abundance Summary\label{tab:tracks}
}
\tablewidth{0pt}
\tablehead{\colhead{Atomic No}&
\colhead{[X/H]} & \colhead{Hyades\tablenotemark{*}} & \colhead{$\sigma$} &
\colhead{Collinder 261} & \colhead{$\sigma$} &
\colhead{HR 1614} & \colhead{$\sigma$} & \colhead{Adopted Solar}
}

\startdata

26 & Fe & 0.13   & 0.05   &-0.01  & 0.02  & 0.24 & 0.03 & 7.52\\
11 & Na & 0.01   & 0.06   & 0.12  & 0.08  & 0.20 & 0.08 & 6.33 \\
12 & Mg & -0.06  & 0.04   & 0.13  & 0.07  & 0.25 & 0.06 & 7.58 \\
14 & Si & 0.05   & 0.04   & 0.18  & 0.05  & 0.27 & 0.05 & 7.55 \\
20 & Ca & 0.07   & 0.07   & 0.00  & 0.09  & 0.24 & 0.05 & 6.36 \\
25 & Mn &\nodata &\nodata &-0.04  & 0.03  & 0.36 & 0.03 & 5.39 \\
28 & Ni &\nodata &\nodata &-0.04  & 0.05  & 0.34 & 0.05 & 6.25 \\
40 & Zr & 0.07   & 0.06   &-0.04  & 0.10  & 0.16 & 0.07 & 2.60 \\
56 & Ba & 0.50   & 0.05   & 0.01  & 0.03  & 0.43 & 0.05 & 2.13 \\
58 & Ce & 0.17   & 0.03   &\nodata&\nodata& 0.13 & 0.03 & 1.58 \\
60 & Nd & 0.01   & 0.03   &\nodata&\nodata& 0.11 & 0.02 & 1.45 \\
63 & Eu &\nodata &\nodata &\nodata&\nodata& 0.21 & 0.03 & 0.52 \\

\enddata
\tablenotetext{*}{Note the Hyades $\alpha$ element abundances were adopted from \citet{P03}.}
\end{deluxetable}

\end{document}